\documentclass[12pt]{article}
\pdfoutput=1
\usepackage{amsmath,amssymb}
\usepackage{graphicx}
\usepackage{cite,fancyhdr}

\newcommand{\Slash}[1]{{\ooalign{\hfil#1\hfil\crcr\raise.167ex\hbox{/}}}}

\newcommand{\beq}{\begin{equation}}  \newcommand{\eeq}{\end{equation}}
\newcommand{\bef}{\begin{figure}}  \newcommand{\eef}{\end{figure}}
\newcommand{\bec}{\begin{center}}  \newcommand{\eec}{\end{center}}
  
\newcommand{\laq}[1]{\label{eq:#1}}  

\newcommand{\Eq}[1]{Eq.(\ref{eq:#1})}
\newcommand{\Eqs}[1]{Eqs.(\ref{eq:#1})}
\newcommand{\eq}[1]{(\ref{eq:#1})}
\newcommand{\Sec}[1]{Sec.\ref{chap:#1}}
\newcommand{\ab}[1]{\left|{#1}\right|}

\newcommand{\lac}[1]{\label{chap:#1}}
\newcommand{\SU}[1]{{\rm SU{#1} } }
\newcommand{\SO}[1]{{\rm SO{#1}} }
\def\({\left(}
\def\){\right)}

\def\O{\mathcal{O}}
\def\U{\mathop{\rm U}}

\def\tr{\mathop{\rm tr}}

\newcommand{\OR}{~{\rm or}~}
\newcommand{\AND}{~{\rm and}~}

\newcommand{\GEV}{ {\rm ~GeV} }
\newcommand{\TEV}{ {\rm ~TeV} }
\def\o{\over}
\def\a{\alpha}
\def\b{\beta}

\def\d{\delta}
\def\e{\epsilon}
\def\f{\phi}

\def\m{\mu}
\def\n{\nu}

\def\s{\sigma}
\def\t{\tau}

\def\D{\Delta}

\def\ol{\overline}
\def\tl{\tilde}
\def\*{\dagger}
\usepackage[affil-it]{authblk}
\usepackage[left=2cm,right=2cm,top=2cm,bottom=2cm,a4paper]{geometry}

\begin{document}

\setcounter{footnote}{0}
\setcounter{figure}{0}
\setcounter{table}{0}

\title{\bf \large 
Flavor-Safe Light Squarks in Higgs-Anomaly Mediation
 }
\author[1,2]{{\normalsize  Tsutomu T. Yanagida}}
\author[3]{{\normalsize Wen Yin}}
\author[4]{{\normalsize Norimi Yokozaki}}

\affil[1]{\small 
Kavli Institute for the Physics and Mathematics of the Universe (WPI),
 
University of Tokyo, Kashiwa 277--8583, Japan}

\affil[2]{\small Hamamatsu Professor}

\affil[3]{\small 
Institute of High Energy Physics,

 Chinese Academy of Sciences, Beijing 100049,  China}

\affil[4]{\small 
Department of Physics, Tohoku University,  

Sendai, Miyagi 980-8578, Japan}

\date{}

\maketitle

\thispagestyle{fancy}
\rhead{ IPMU18-0016 \\ TU-1056 }
\cfoot{\thepage}
\renewcommand{\headrulewidth}{0pt}

\begin{abstract}
\noindent
We consider a simple setup with light squarks which is free from the gravitino and SUSY flavor problems. 
In our setup, a SUSY breaking sector is sequestered from the matter and gauge sectors, and it only couples to the Higgs sector directly with $\O(100)$\,TeV gravitino. Resulting mass spectra of sfermions are split: the first and second generation sfermions are light as $\mathcal{O}(1)$\,TeV while the third generation sfermions are heavy as $\mathcal{O}(10)$\,TeV. 
The light squarks of $\mathcal{O}(1)$\,TeV can be searched at the (high-luminosity) LHC and future collider experiments. 
Our scenario can naturally avoid too large flavor-changing neutral currents and it is consistent with the $\e_K$ constraint.
Moreover, there are regions explaining the muon $g-2$ anomaly and
bottom-tau/top-bottom-tau Yukawa coupling unification simultaneously.
\end{abstract}

\section{Introduction}

The minimal supersymmetric (SUSY) extension of the standard model (MSSM) has been known as a promising candidate of the particle physics beyond the standard model (SM). This is because it goes well with the grand unified theory explaining the hypercharge quantization, has interesting dark matter candidates, and relaxes the 
hierarchy problem between the fundamental and the electroweak (EW) scales.
However, the MSSM also brings difficulties in general, such as SUSY flavor-changing neutral current (FCNC) problem and gravitino problem, 
which should be solved in a realistic setup.

 It was pointed out that if SUSY breaking is mediated by gravity, the flavor-violating parameters of the MSSM 
 would be induced by the mediation of Planck-scale states. 
This problem was solved by ``sequestering" the MSSM sector from the SUSY breaking sector~\cite{Inoue:1991rk}. 
 In the setup, gaugino and sfermion masses are generated through anomaly mediation~\cite{Giudice:1998xp, Randall:1998uk}.  However, the simplest possibility suffers from the so-called tachyonic slepton problem: mass squared parameters of sleptons are negative. 

Recently the tachyonic slepton problem is solved in Higgs-Anomaly mediation~\cite{Yin:2016shg, Yanagida:2016kag},\footnote{
The Higgs-Anomaly mediation might be regarded as a variant of pure gravity mediation~\cite{Ibe:2006de,Ibe:2011aa} or minimal split SUSY~\cite{ArkaniHamed:2012gw}, where the SUSY breaking effects are purely mediated to the visible sector by gravity with $\O(10$\,-\,$100)\TEV$ gravitino.
} where only Higgs doublets couple to the SUSY breaking sector through Planck-suppressed operators while the matter and gauge sectors are sequestered from it as in the setup of Ref.~\cite{Randall:1998uk}. 
The sequestering might be due to the separation between the SUSY breaking and matter branes in an extra-dimension setup, or because squarks and sleptons are pseudo Nambu-Goldstone bosons of $E_7/{\SU(5)\times \U(1)^3}$~\cite{Yanagida:2016kag}. 
In this setup, gaugino masses are determined by anomaly mediation while the sfermion masses get additional contributions from Higgs-loops. As a result, the slepton mass squared parameters are positive at a low-energy scale. The stop mass scale is $\mathcal{O}(10)$\,TeV with $\mathcal{O}(100)$\,TeV gravitino. This stop mass scale can naturally explain the measured Higgs boson mass of 125\,GeV, taking into account radiative corrections~\cite{Okada:1990vk,  Ellis:1990nz,  Haber:1990aw, Okada:1990gg, Ellis:1991zd}. Moreover, the gravitino problem is solved due to the earlier decay than the epoch of big bang nucleosynthesis, thanks to the heavy gravitino~\cite{Kawasaki:2008qe}.

In Higgs-Anomaly mediation, all the slepton and squark masses vanish at the tree-level, and thus there is no flavor-violation at this level. However, splitting mass spectra are generated at quantum level dominantly through Higgs-loop effects~\cite{Yamaguchi:2016oqz}: the sfermion masses are hierarchical in generations at the low-energy scale. This hierarchy originates from the hierarchy of the SM Yukawa couplings, and we have light sleptons and squarks in the first two generations which can be searched at the LHC and future collider experiments~\cite{fc,fc1,fc2,fc3,fc4}.
Surprisingly, this simple setup can explain the muon magnetic dipole moment ($g-2$) anomaly and bottom-tau/top-bottom-tau Yukawa unification. In particular, there is a robust prediction: if the muon $g-2$ anomaly is explained at the $1\s$ level, the masses of the light squarks are fully within the sensitivity reach of the LHC.

However, it is not clear whether such light squarks are safe from the SUSY FCNC problem.
In this paper, a precise analysis is performed to show that the problem is solved in Higgs-Anomaly 
mediation, focusing on the $K$-meson mixing. 
In the previous paper~\cite{Yin:2016shg}, the estimation of the $K$-meson mixing was based on the mass insertion approximation, and it was found to be marginal to the experimental constraint~\cite{Patrignani:2016xqp} in the muon $g-2$ region. 
Thus we need to check whether the FCNC processes are safe without the mass insertion approximation regarding current constraints and lattice QCD results \cite{Aoki:2016frl,Jang:2017ieg}.

Then, we revisit the SUSY spectrum of the Higgs-Anomaly mediation with a precise analysis taking account of the off-diagonal elements of the mass matrices and the renormalization scale of the threshold corrections for dimensionless couplings. We show that the region explaining the muon $g-2$ anomaly is enlarged, while the region consistent with the Yukawa coupling unification moves toward the muon $g-2$ region. 
As a result, we find some overlapped regions.

This paper is organized as follows. In \Sec{2} we explain the Higgs-Anomaly Mediation and show the typical squark spectra with splitting mass.
In \Sec{3} the flavor safety of the whole viable region is discussed. In \Sec{pred} we show the viable region explaining the muon $g-2$ anomaly or/and the Yukawa coupling unifications. The last section is devoted to conclusions.

\section{Higgs-Anomaly Mediation and Light Squarks}
\lac{2}

Let us explain the setup of Higgs-Anomaly mediation. 
In Higgs-Anomaly mediation, the Higgs soft masses are non-vanishing and negative. The sfermion masses are generated through anomaly mediation and Higgs-loop effects, while gaugino masses are simply determined by anomaly mediation. 

At $M_{\rm inp}=10^{16}$\,GeV, we take the tree-level mass parameters as  
\begin{equation}
m_{H_u}^2=m_{H_d}^2=c_h m_{3/2}^2<0,~ \m=\m_0, ~B=B_0,
\end{equation}
\begin{equation}
\laq{gaugino}
M_1=M_2=M_3=0,\end{equation}
\begin{equation}
~ {\bf A}_u={\bf A}_d={\bf A}_e=0,~\end{equation}\begin{equation}
{\bf m}_{\tl{ Q}}^2={\bf m}_{\tl{ u}}^2={\bf m}_{\tl{ d}}^2={\bf m}_{\tl{ L}}^2={\bf m}_{\tl{ e}}^2=0,
\end{equation}
where $m_{H_u}$ and $m_{H_d}$ are up-type and down-type Higgs soft masses, respectively; $M_1$, $M_2$ and $M_3$ are bino, wino and gluino masses, respectively; ${\bf A}_{u,d,e}$ is a matrix of scalar trilinear couplings and ${\bf m}_{\tilde X}$ is a sfermion mass matrix of a field $\tilde X$. Both the matrices are $3\times 3$ in generations.
Here $c_h$ represents the Higgs soft masses in the unit of gravitino mass, $m_{3/2}$. We consider $c_{h}<0$, which is crucial to solve the tachyonic slepton problem and get large stop masses of $\mathcal{O}(10)$\,TeV.
In our setup, a CP symmetry in the SUSY breaking sector is assumed so that $B_0$ and $m_{3/2}$ are both taken to be real without loss of generality. Throughout this paper, we consider $\m_0>0$ for simplicity, 
while the discussions hereafter hold with $\m_0<0$, unless otherwise stated.

As briefly mentioned, the sfermion and gaugino masses are raised by quantum corrections. 
The sfermion masses are dominantly composed of two sources of radiative corrections\footnote{In a five-dimensional spacetime scenario, the Higgs doublets may live in the bulk with size $L$, and couple to the matters and SUSY breaking field with higher dimensional terms on each brane. Then, there is a bulk-Higgs one-loop contribution to the sfermion masses of order ${1\over 16\pi^2} {1\over L^2  M_{\rm pl}^2}c_h m_{3/2}^2= \O({10^{-8}}) \({1/L \over 10^{16}\GEV}\)^2 \({c_h\over -0.1}\) m_{3/2}^2$ (c.f. Ref. \cite{Rattazzi:2003rj}) breaking the sequestering. The masses may have flavor-violating components but is negligible for FCNC processes (see Ref.\cite{Gabbiani:1996hi, Altmannshofer:2009ne} and \Sec{3}), if $1/L\sim 10^{16}\GEV$. To obtain $y_t,y_b,y_\tau \sim 0.5$ at $M_{\rm inp}$, the Yukawa couplings in five-dimension with fundamental scale of $\sim (M_{\rm pl}^2/L)^{1/3}$ are of $\O(1)$, which can be further reduced by taking the bulk size slightly smaller. We thank H. Murayama for discussion on this flavor-violating effect.}\footnote{ In the scenario that the sfermions are pseudo Nambu-Goldston bosons, we can consider a non-compact K\"{a}hler manifold ${\cal K}_{E_7}$ \cite{Kugo:2010fs} for the non-linear sigma model with an $E_7$ breaking scale $\O(10^{16}\GEV)$. This model can couple to supergravity which is broken in a hidden sector \cite{Goto:1990me}. 
The resulting effective K\"{a}hlar potential, ${\cal K}_{E_7}+\ab{Z}^2(1-c_h {1\over M_{\rm pl}^2}(|H_u|^2+ |H_d|^2)...)$, does not have flavor-violating parameters. 
Since the explicit breaking of $E_7$ only comes from the Yukawa and gauge couplings, there are no flavor-violating sources other than the Yukawa couplings.
 }: anomaly mediation, $\d^{\rm AM}{\bf m}_{\tl{ X}}^2$, and Higgs loop effects, $\d^{\rm HM}{\bf m}_{\tl{X}}$.
The squark masses are 
\begin{equation}
{\bf m}_{\tl{ Q},\tl{ u},\tl{ d}}^2=\d^{\rm AM}{\bf m}_{\tl{ Q},\tl{ u},\tl{ d}}^2+\d^{\rm HM}{\bf m}_{\tl{ Q},\tl{ u},\tl{ d}}^2.
\end{equation}
For instance, anomaly mediation gives a squark mass squared at two-loop level as~\cite{Randall:1998uk} 
\begin{equation}
\laq{Anmd}
\d^{\rm AM} {\bf m}_{\tl{Q},\tl{u},\tl{d}}^2 \simeq {\bf 1} \({g_3^2 \o 16\pi^2}\)^2 8 m_{3/2}^2 .
\end{equation}

\paragraph{Higgs Mediation}

The contribution from the renormalization group (RG) running effects from Higgs loops is a key feature of our setup. We call it Higgs mediation. 
For illustrative purpose, we show the contribution to squark masses with the leading log approximation: 
\begin{align}
\laq{mQ}
\d^{\rm HM}_{\rm 1loop} {\bf m}_{{\tl{Q}}}^2 &\simeq {2 \o 16 \pi^2}  \({\bf Y}_u^\* \cdot {\bf Y}_u  +{\bf Y}_d^\* \cdot {\bf Y}_d\) c_h m_{3/2}^2\log{\({m_{\tl{t}}\o M_{\rm inp}}\)} ,\\
\laq{mu}
\d^{\rm HM}_{\rm 1loop} {\bf m}_{{\tl{u}}}^2 &\simeq {4 \o 16 \pi^2} \({\bf Y}_u \cdot {\bf Y}_u^\* \) c_h m_{3/2}^2\log{\({m_{\tl{t}}\o M_{\rm inp}}\)} ,\\
\laq{md}
\d^{\rm HM}_{\rm 1loop} {\bf m}_{{\tl{d}}}^2 &\simeq {4 \o 16 \pi^2}  \({\bf Y}_d \cdot {\bf Y}_d^\* \) c_h m_{3/2}^2\log{\({m_{\tl{t}}\o M_{\rm inp}}\)},
\end{align}
where ${\bf Y}_u \AND {\bf Y}_d$ are the Yukawa matrices for up and down-type quarks, respectively.
They are given by
\begin{equation}
{\bf Y}_u= \left(\begin{array}{ccc}y_u & 0 & 0 \\0 & y_c & 0 \\0 & 0 & y_t \end{array}\right) \cdot V,\AND {\bf Y}_d=\left(\begin{array}{ccc}y_d & 0 & 0 \\0 & y_{s} & 0 \\0 & 0 & y_b \end{array}\right),
\end{equation}
using the CKM matrix $V$.  The MSSM Yukawa couplings are matched to the SM ones:
\begin{equation} 
\laq{yukawas}
y_{U} \simeq {{m_{U}\o v} {\sqrt{1+\tan^2\b}\o \tan\b}} \AND y_{D} \simeq {m_{D}\o v}{ \sqrt{1+
\tan^2\b} \o 1+\Delta_D} , \end{equation} where 
\begin{equation}\laq{th} \D_D \simeq {g_3^2\o 6\pi^2} M_3 \m \tan\b I(m_{\tl{D}_L}^2,m_{\tl{D}_R}
^2,M_3^2) . \end{equation}
The index $U$ ($D$) denotes $\{u,c,t\}$ ($\{d,s,b\}$), and $v\simeq 174\GEV$ is the Higgs vacuum expectation value;
$\tan\beta$ is a ratio of the vacuum expectation values, $\left<H^0_u\right>/\left<H^0_d\right>$;
$I(x,y,z)$ is a loop function given by
\begin{equation} I(x,y,z) = -\frac{xy\ln (x/y) + yz \ln (y/z) + zx \ln (z/x)}{(x-y)(y-z)(z-x)}.\end{equation} 
Note that $\Delta_D$ dominantly comes from the threshold correction via a gluino-squark loop. 

As one can see from Eq.\,(\ref{eq:mQ}) to (\ref{eq:md}), the squark masses are split depending on the Yukawa couplings: the masses of the stops are $\O(10)\TEV$ for $c_h m_{3/2}^2\sim -(100\TEV)^2$ while those of the first and second generation squarks are much smaller. We note that the sbottoms and staus are also heavy as $\mathcal{O}(10)$\,TeV due to the large $\tan\beta$ as described below.

\paragraph{Splitting Mass Spectra and the EW scale}
In our setup, $m_{H_u}^2$ and $m_{H_d}^2$ are negative and large.
Therefore, one needs a large $\mu$-term to lift the $D$-flat direction: the coefficients of the Higgs quadratic terms need to be positive at the tree-level, i.e. $c_h m_{3/2}^2 + \mu_0^2 \gtrsim 0$.

Taking into account the radiative corrections, large $y_b, y_\t \sim \O(y_t)$ are also required for a successful EW symmetry breaking (EWSB). In order to have a positive mass squared for the CP-odd Higgs, 
$m_{H_u}^2 + m_{H_d}^2 + 2 \mu^2 \simeq m_{H_d}^2 - m_{H_u}^2 > 0$ is required at the stop mass scale. 
The Higgs soft masses at the stop mass scale are determined by the RG equations (RGEs), which are given by
\begin{equation}
{d m_{H_u}^2\over d \ln\m_{R}} \supset {1\o16\pi^2} 6 y_t^2 m_{H_u}^2, ~{d m_{H_d}^2\over d \ln\m_{R}} \supset {1\o16\pi^2}\({6 y_b^2+ y_\t^2}\) m_{H_d}^2,
\end{equation}
where $\m_R$ is the renormalization scale. 
Therefore $m_{H_d}^2 > m_{H_u}^2$ is achieved for $y_b \AND y_\t$ of the order of $y_t$. (Remember that $m_{H_u}^2$ and $m_{H_d}^2$ are both negative.) The condition for the Yukawa couplings leads to the large $\tan\beta$, $\tan\b \gtrsim 45$.
Then, the sbottom and stau masses are also raised through Higgs mediation [see. Eq.(\ref{eq:mQ})-(\ref{eq:md})]: all the third generation sfermions have masses of order $\O(0.1) m_{3/2}$ for $c_h=-\O(1)$, 
while the first two generation sfermions have much smaller ones.\footnote{Since the heavy sfermions can be embedded into GUT multiplets, we have gauge coupling unification.}

\begin{figure}[!t]
\begin{center}  
   \includegraphics[width=80mm]{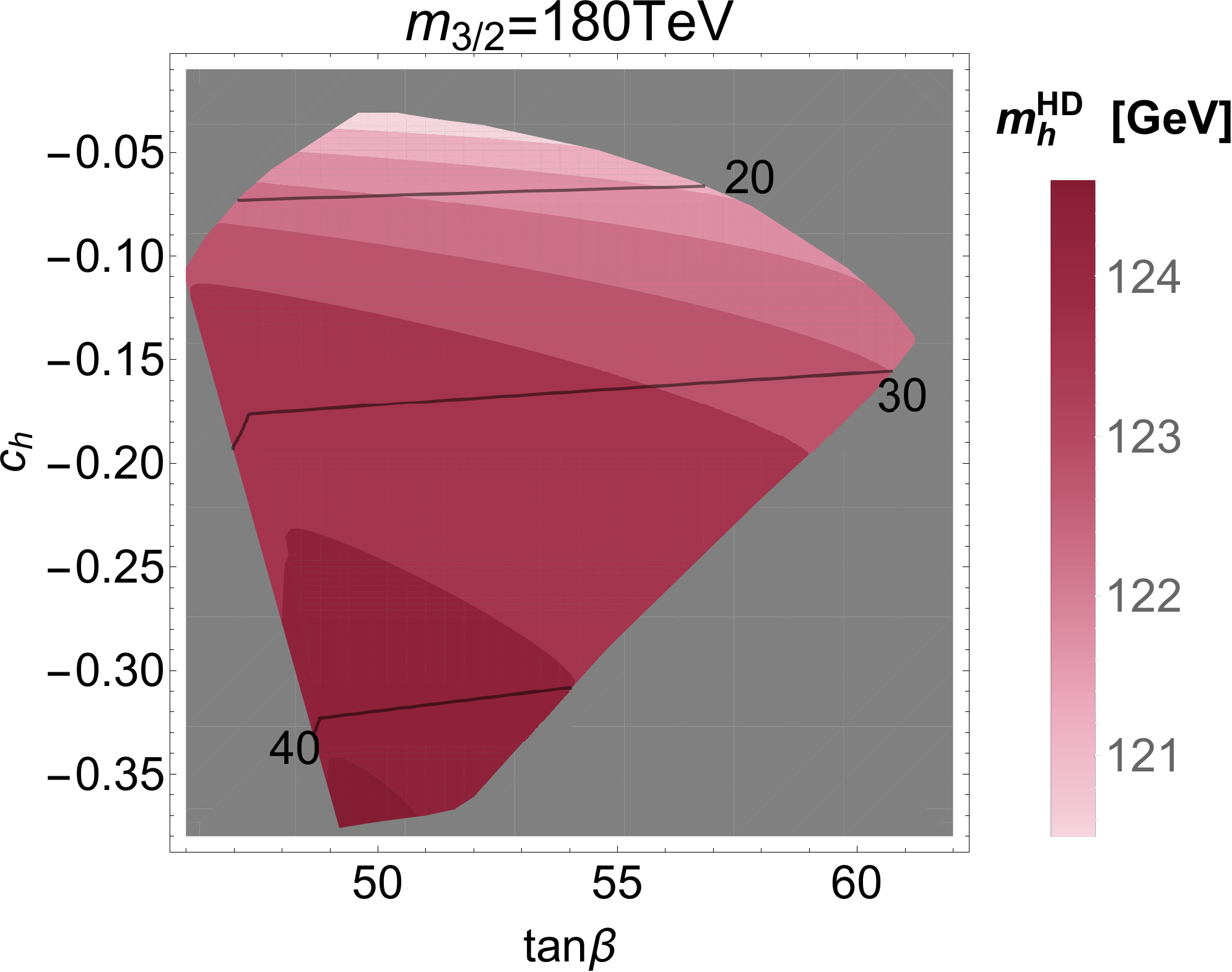}
   \includegraphics[width=80mm]{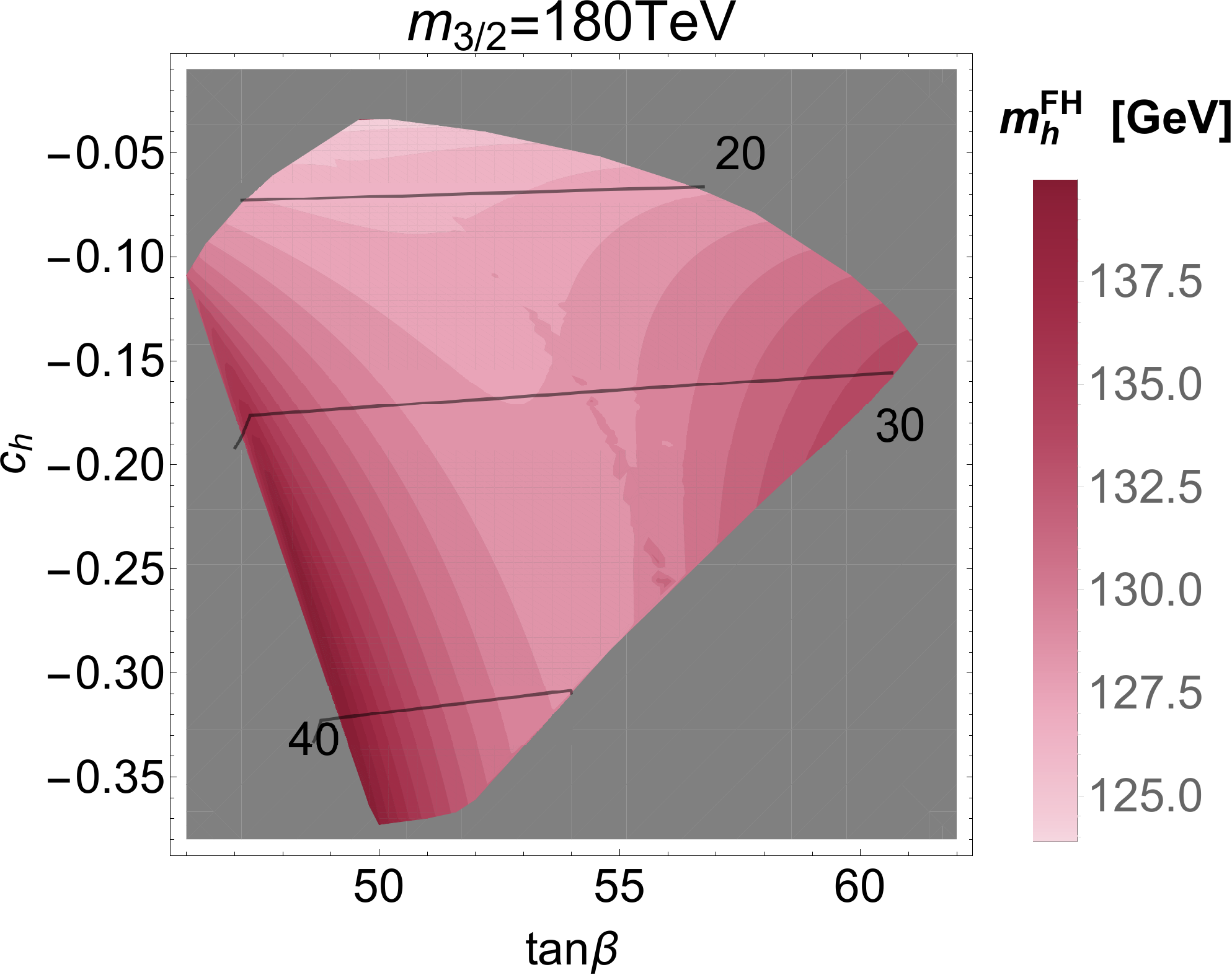}
\end{center}
\caption{Contour plots of the Higgs boson mass [GeV] calculated using {\tt SUSYHD} (left) and  {\tt FeynHiggs} (right). 
We take $m_{3/2}=180\,\TEV$. The stop mass scale is also shown in units of TeV.
{ Here, $\alpha_s(m_Z)=0.1185$ and $m_t({\rm pole})=173.34$\,GeV.}}
\label{fig:Higgs}
\end{figure}

The Higgs boson and the stop mass scale $\sqrt{m_{\tl{t}_L}m_{\tl{t}_R}}$ 
are shown in Fig.\ref{fig:Higgs}. The SUSY mass spectra are computed using {\tt SuSpect\,2.4.3}~\cite{Djouadi:2002ze} with 
appropriate modifications. 
In particular, we have modified the single matching scale for the dimensionless couplings in the {\tt SuSpect} into two scales: the stop mass scale and gluino mass scale (c.f. Ref. \cite{Chigusa:2016ody, Chigusa:2017drd}). 
{ In between the SM and gluino mass scales the dimensionless couplings are obtained by solving SM RGEs at two-loop level~ \cite{Buttazzo:2013uya}, and they are obtained by solving the RGEs of SM + gauginos at the two-loop level~\cite{Giudice:2011cg} in between the gluino and stop mass scales. Then, the SUSY threshold corrections are added at the stop mass scale~\cite{Pierce:1996zz}.}
This modification reduces $y_b$ by around $10\%$ for a given $\tan \b$ at the stop mass scale compared with the previous works (see \Sec{pred})~\cite{Yin:2016shg, Yanagida:2016kag}.

The Higgs boson masses estimated using {\tt SUSYHD 1.0.2}~\cite{Vega:2015fna} and {\tt FeynHiggs 2.13.0}~\cite{feynhiggs,feynhiggs2,feynhiggs3,feynhiggs4,feynhiggs5,feynhiggs6,feynhiggs7} are shown in the left and right panels, respectively.
The Higgs boson mass increases with larger $\ab{c_h}$ for fixed $\tan\b$ in both codes 
because the stop mass scale increases due to the Higgs mediation.  
Given $c_h\AND \tan\b$, the Higgs boson mass also increases with larger $m_{3/2}$ for the same reason. 
There are large discrepancies between these two results in the calculations.\footnote{ {\tt SUSYHD} ({\tt FeynHiggs}) purely (partially) performs an effective field theory calculation, where the resummation of large logarithm terms are made. Thus large logarithm terms, which are partially not resummed in the {\tt FeynHiggs}, might be the dominant discrepancy at large $|c_h|$. 
On the other hand, the threshold correction to the Higgs boson mass of order $\({1\over 16\pi^2}\)^2{ y_b^6 \m^6 \o m_{\tl{b}}^6}v^2$ is not included in {\tt SUSYHD} while is included in {\tt FeynHiggs}. Due to the $\m \tan\b$ enhancement, this term can contribute several GeVs to the Higgs boson mass, for small $\ab{c_h}$ and large $\tan\b$.  }   Also, there might be several GeVs uncertainties in the calculations due to higher order missing terms (cf. Ref.~\cite{Draper:2016pys,Athron:2016fuq,Bagnaschi:2017xid}).
Therefore, within the uncertainty, we conservatively consider the regions discussed here and hereafter can explain the observed Higgs boson mass.\footnote{For example, the threshold correction terms of order $ \({1\over 16\pi^2}\)^2 \({\m^2  \over m_{\tl{b}/\tl{\tau}}^2}\)^3 v^2 \times (y_b^4 y_\tau^2, y_b^2 y_\tau^4, y_\tau^6 )$ are not included in both codes.  We have checked the Higgs boson mass evaluated by {\tt FlexibleSUSY 2.1.0}~\cite{Athron:2017fvs, Athron:2016fuq}, where some of the missing two-loop threshold corrections are included.
The result is more consistent with ${\tt SUSYHD}$, but there is $\O(1)\GEV$ increase of the Higgs boson mass. 
The increase can be up to $2-3\GEV$ in the region with $\ab{c_h}=\O(0.01)$ and large $\tan\b$.}

\paragraph{Light Squarks} \lac{lsq}
The light squarks in the first two generations appear as the interesting features of our scenario. 
In particular, the masses of the right-handed squarks decrease during the RG running at the two-loop level. 
The contribution from the RG running is proportional to the mass squared parameters of the third generation squarks as
\begin{equation}
{d {\bf m}_{\tl{Q},\tl{u},\tl{d}}^2\over d \ln\m_{R}}\supset{\bf 1}\times  \({g_3^2\o 16\pi^2}\)^2{16\o 3}   {\rm Tr}{[2{\bf m}_{\tl{Q}}^2+{\bf m}_{\tl{u}}^2+{\bf m}_{\tl{d}}^2]}. \label{eq:mqud}
\end{equation}
Since the masses of the third generation squarks are positive and $\mathcal{O}(0.1)\,m_{3/2}$ at the low-energy scale, 
the above contribution is negative. Thus the masses of the first two generation squarks are reduced.

On the other hand, there is a positive contribution to ${\bf m}_{\tl{Q}}^2$. It is raised via the two-loop effects through $\SU(2)$ gauge interaction, 
\begin{equation}
\d^{\rm HM,2loop} {\bf m}_{{\tl{Q}}}^2 \simeq {\bf 1}\times  \({g_2^2 \o 16 \pi^2}\)^2 6 c_h m_{3/2}^2 \log{\({m_{\rm \tl{t}} \o M_{\rm inp}}\)}, 
\end{equation} 
where we use the leading log approximation. The positive effect is comparable to the one in Eq.(\ref{eq:mqud}).
To sum up, the diagonal components of squark mass squared parameters in the first two generations satisfy, 
\begin{equation}
{\bf m}_{\tl{Q},ii}^2 \sim \d^{\rm AM}{\bf m}_{\tl{ Q},ii}^2,\ 
{\bf m}_{\tl{u}/\tl{d},ii}^2\lesssim \d^{\rm AM}{\bf m}_{\tl{ u}/\tl{d},ii}^2,
\end{equation}
where the index $i$ takes $d \OR s$. The right-handed squarks are always lighter than gluino mass determined by anomaly mediation~\cite{Randall:1998uk, Giudice:1998xp}
\begin{equation}
\laq{gl}
M_{3} = -3 {g_3^2 \o 16\pi^2}m_{3/2}.
\end{equation}
The masses of the bino and wino are given by
\begin{eqnarray}
\laq{w}
M_{1} = \frac{33}{5} {g_1^2\over 16\pi^2} m_{3/2}, \AND M_{2} ={ g_2^2\over 16\pi^2} m_{3/2}, 
\end{eqnarray}
respectively. 
Notice that since $\tan\b=\O(10)$ in our setup, the threshold correction from the Higgs-higgsino loop is negligible~\cite{Pierce:1996zz,Yanagida:2016kag}.

\begin{figure}[!t]
\begin{center}  
   \includegraphics[width=75mm]{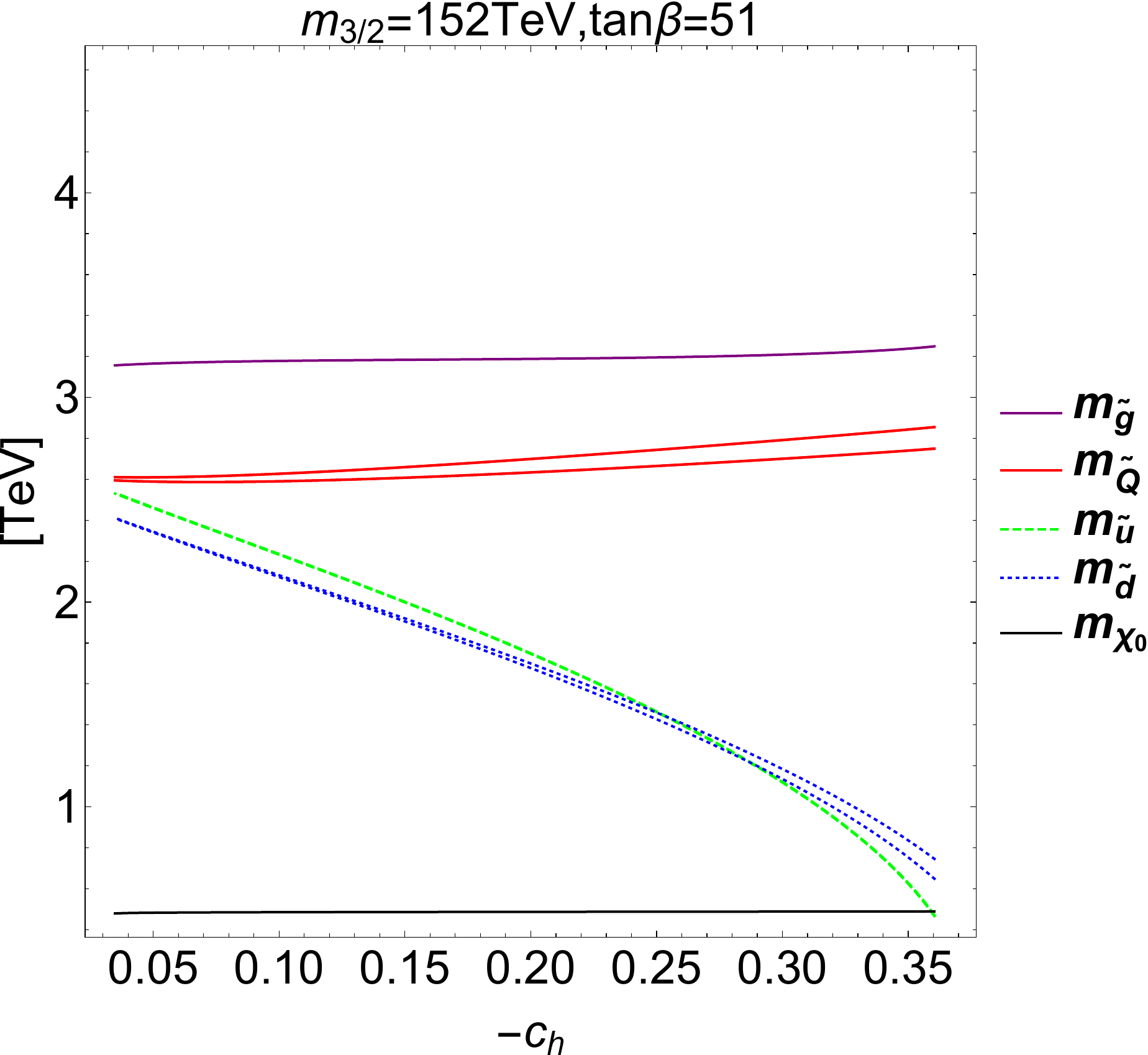}
   \includegraphics[width=75mm]{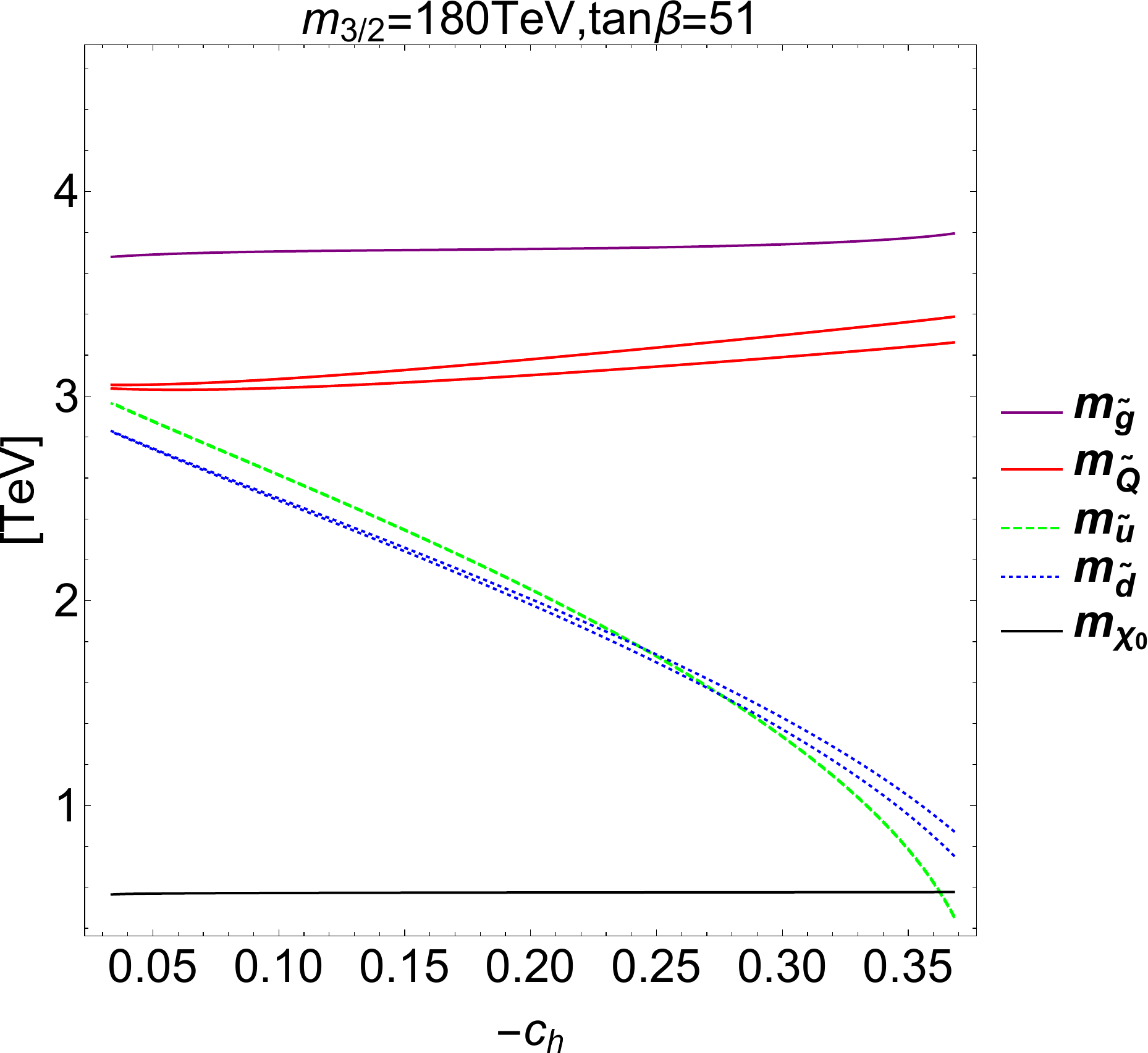}
\end{center}
\caption{ The mass spectra of colored sparticles. 
The gravitino mass and $\tan\b$ are fixed as $m_{3/2}=152\TEV$ and $\tan\b=51$ ($m_{3/2}=180\TEV$ and $\tan\b=51$) in the left (right) panel. The black solid line denotes the wino mass for comparison.}
\label{fig:sq}
\end{figure}

The dependence of the first two generation squark masses on $c_h$ are shown in Fig.\ref{fig:sq} for fixed $\tan\b$ and $m_{3/2}$.
In the numerical calculation, we have taken account of the off-diagonal components of the mass matrices and the Yukawa couplings for the first two generations. 
Notice that the masses of the right-handed up-type squarks are almost degenerate due to the suppressed Yukawa couplings of charm and up quarks. (The left-right mixing is negligibly small.)

This system is severely constrained for a small $|c_h|$ in the LHC due to the large production rate of light colored sparticles~\cite{Aaboud:2017vwy}. {The squarks in this region will be tested up to $\sim$ 3~TeV~\cite{Cohen:2013xda}}, corresponding to $m_{3/2}\lesssim 200\TEV$. 
Moreover, the squarks as well as the gluino with the masses up to around 7\,TeV and 15\,TeV are within the projected sensitivity of the 33\,TeV and 100\,TeV colliders~\cite{Cohen:2013xda, Arkani-Hamed:2015vfh, Golling:2016gvc}. 
The corresponding gravitino masses are $m_{3/2}\simeq 400\,\TEV$ and $1000\,\TEV$, respectively.

In the region with larger $|c_h|$ some of the squark masses are comparable to the wino mass. (We focus on the region where the wino is the lightest SUSY particle.) 
In this region, the LHC constraints are relaxed because a jet produced from the squark decay has a too small transverse energy. The LHC can test light squarks with wino mass $\lesssim 600\GEV$, corresponding to $m_{3/2}\lesssim 200\TEV$\cite{Cohen:2013xda, Arkani-Hamed:2015vfh, Golling:2016gvc}.
This compressed region can be tested in the 33\,TeV and 100\,TeV colliders with the wino mass up to $1.2\TEV$ and 4\,TeV, equivalently $m_{3/2}\simeq 400\TEV$ and 1300\,TeV.

In fact, a pure wino dark matter of mass smaller than $3\TEV$ ($m_{3/2} \lesssim 1$\, PeV) can explain the observed dark matter abundance with non-thermal production through gravitino decay.\footnote{Strictly speaking, when $|c_h|$ is sufficiently large, one should take into account the coannihilations between the wino and light squarks due to the compressed spectrum. This could enlarge the parameter region (c.f. Ref. \cite{Hisano:2006nn, Ellis:2015vaa}).} The pure wino dark matter is viable even with current AMS-02 antiproton constraints~(c.f. Ref. \cite{Reinert:2017aga}).\footnote{We thank S. Matsumoto for useful communication on this issue.}  
In particular, when the mass of the wino is smaller than $1\TEV$ ($m_{3/2}\lesssim 300\TEV$), the reheating temperature of the universe can be high enough for thermal leptogenesis~\cite{Fukugita:1986hr} (see also \cite{Buchmuller:2005eh, Davidson:2008bu} for reviews) without producing too much dark matter through gravitino decay \cite{Yanagida:2016kag}. This region is within the reach of the 33\,TeV collider.

\section{Flavor Safety}

\lac{3}

The Higgs mediation effect \eq{mQ} has flavor-dependence.
Thus it is quite non-trivial whether flavor-violating processes are suppressed to be consistent with the current measurement \cite{Patrignani:2016xqp}, especially with light squarks. 
Suppressions of flavor-violating processes in the lepton sector are guaranteed
 because the lepton Yukawa matrix can be diagonalized.
The only flavor-violating sources are the quark Yukawa couplings. 
The flavor-violating contribution from SUSY appears through these Yukawa couplings via the left-handed squark mass squares induced by the Higgs mediation and via squark left-right mixing terms.
This is categorized into the so-called minimal flavor violation scenario \cite{Hall:1990ac,
Ciuchini:1998xy,
Buras:2000dm,
DAmbrosio:2002vsn, Paradisi:2008qh}.\footnote{Notice that our scenario is not within the 
universality-class of the fixed-point discussed in Ref. \cite{Paradisi:2008qh,Colangelo:2008qp}.
This is because the gauge contributions to the squark masses in our scenario are suppressed compared with the Higgs mediation, while the gauge contribution was relevant for the fixed-point. } 
Since the stop and sbottom in most of the region are heavy as $m_{\tl{b},\tl{t}} = \O({0.1}
m_{3/2})$,  flavor-violating processes in $B$-meson system are negligible.\footnote{There are also tiny 
regions with $m_{A}= \O(100)\GEV$ where we may have significant contributions to the anomalous decays of mesons (cf. \cite{Freitas:2007dp}).} 

The dominant FCNC process is in $K$-meson system due to the light squarks in the first two generations.
A na\"{i}ve mass insertion approximation~\cite{Gabbiani:1996hi, Altmannshofer:2009ne} 
 is no longer valid due to the splitting spectrum of the squarks. 
This is the reason why we carry out precise numerical calculations in the mass eigenbasis. The results of the calculations will be shown later.
 
Before the precise calculations, let us analytically discuss the
$K^0$\,-\,$\ol{K^0}$ meson mixing
with the approximation of the effective mass insertion for illustrative purpose. In the approximation, we integrate out the third generation squarks. 

The $(b,b)$ component of the mass matrix for the left-handed squarks is approximately given by
\begin{equation}
{\bf m}^2_{\tl{Q}b,b} \simeq {1\over 8\pi^2} (\ab{V_{tb}}^2 y_t^2 m_{H_u}^2+y_b^2 m_{H_d}^2)  \log{\({m_{\tl{t}}\o M_{\rm inp}}\)},
\end{equation}
which is much larger than the others, e.g.
\begin{equation}
{\bf m}^2_{\tl{Q}d,b} \simeq {1\over 8\pi^2} ({V_{t d}^* V_{tb}} y_t^2) m_{H_u}^2 \log{\({m_{\tl{t}}\o M_{\rm inp}}\)},
\end{equation}
where $\ab{V_{td}}\simeq 8.7 \times 10^{-3} ,\ab{V_{tb}}\simeq 1$. Here, we have taken the super-CKM basis.
We have neglected the contributions involving gauge couplings and $y_u, y_c, y_d,y_s$.

After integrating out the ($b,b$) component, we obtain a $2\times2$ effective mass matrix
 for the first two generations:
\begin{equation}
{\bf m}_{\tl{Q}i,j}^{2,{\rm \rm eff}}\simeq {\bf m}^2_{\tl{Q}i,j}-{1\over 8\pi^2} \( {y_t^4 m_{H_u}^2 {V_{t i}^*V_{tj} \ab{V_{tb}}^2} \over y_t^2\ab{V_{tb}}^2m_{H_u}^2+y_b^2 m_{H_d}^2}\) m_{H_u}^2 \log{\({m_{\tl{t}}\o M_{\rm inp}}\)},
\end{equation}
where ${\bf m}^2_{\tl{Q}i,j}$ is given in Eq.(\ref{eq:mQ}). 
We see that the flavor-violating off-diagonal component obtains a suppression factor $r \equiv  \( {y_b^2 m_{H_d}^2  \over y_t^2\ab{V_{tb}}^2 m_{H_u}^2+y_b^2 m_{H_d}^2}\)$ as
\begin{equation}
{\bf m}_{\tl{Q}d,s}^{2, {\rm \rm eff}}\simeq r {\bf m}_{\tl{Q}d,s}^{2} \simeq r {1  \over 8\pi^2}  V_{td}^*V_{ts}  y_t^2 m_{H_u}^2 \log{\({m_{\tl{t}}\o M_{\rm inp}}\)}.
\end{equation}
This suppression factor, $r$, reflects the fact that ${\bf Y}_u$ can be diagonalized and all flavor-violating sources disappear in the limit $y_b \rightarrow 0$. %
Notice that even for $y_t\sim y_b$, $r$ is $\O(0.1)$. %
Interestingly, with the splitting mass spectrum, the flavor-violating effect is even suppressed compared to the case of almost degenerate squark masses.

The $(d,d)$ and ($s,s$) components of this $2\times2$ matrix mostly come from gauge contributions, which are almost the same. In oder to use the mass insertion approximation, we define 
\begin{equation}
(\d^{d, {\rm \rm eff}}_{LL})_{12}\simeq {2{\bf m}_{\tl{Q}d,s}^{2,{\rm \rm eff}}\over {\rm tr}[{\bf m}_{\tl{Q}}^{2,{\rm \rm eff}}]} \simeq -(20+7 i)\times 10^{-3} \({V^*_{td}V_{ts} \over (-3- 1i)\times 10^{-4}}\) \({c_h\over -0.2}\) \({y_t \over 0.7}\)^2  \({r\o 0.5}\),
\end{equation}
where we use \Eq{Anmd} to evaluate ${\rm tr}[{\bf m}_{\tl{Q}}^{2,{\rm \rm eff}}]$.  We see that the constraints from $\D M_K$ and $\e_K$ are satisfied for ${\rm tr}[{\bf m}_{\tl{Q}}^{2,{\rm \rm eff}}]/2\gtrsim (2\TEV)^2 $(see Appendix A).

Let us discuss the SUSY contribution to the $K$-meson mixing in more detail.
In our mass range where $ 0.7 M_{3} \lesssim \sqrt{{\bf m}
^{\rm 2, \rm eff}_{\tl{Q},ii}}  \lesssim 0.9 M_3$, there is a cancellation in the gluino contribution, which has the two box diagrams with and without a cross. These diagrams have opposite signs and cancel each other. 
As a result, there is a suppression factor of $r'= \O(0.1)$.

The chargino and neutralino contributions are also dominant since there is no cancellation. 
Note that the gluino-neutralino contribution has diagrams with and without a cross. 
However, these diagram contributes constructively due to $M_3/M_1<0 \AND M_3/M_2 <0$ [see \Eq{mixed}]. The analytic formulae for the dominant contributions to $M_{12}^K$ are shown in Appendix A with the approximation of the effective mass insertion which agrees with the numerical estimation with an error smaller than $20\%$.\footnote{To compare with the numerical result, 
${\bf m}^2_{\tl{Q}}$ is calculated by solving the RGEs, whose $(b,b)$ component is integrated out to lead to ${\bf m}
^{\rm 2, \rm eff}_{\tl{Q},ii}$. } 
To sum up, compared with the na\"{i}ve mass insertion approximation without cancellation of box diagrams, there is a suppression factor $r^2r'= \O(10)\%$.

Now, we numerically calculate $M_{12}^K$ in the mass eigenbasis taking into account the left-right mixing as well as other sub-dominant contributions. The formulae of the SUSY contribution can be found in Ref. \cite{Goto}. {In the calculation, we solve one-loop RGEs for the off-diagonal components of the squark mass matrix for the Higgs mediation 
by following Ref.\cite{Paradisi:2008qh} without neglecting $y_u, y_c, y_d, y_s$ given in \Eqs{yukawas}.}
The diagonal and off-diagonal masses of the squarks from anomaly mediation are also added at the stop mass scale. 

Let us focus on $\e_K$, which is precisely measured as \cite{Patrignani:2016xqp}
\begin{equation}
\ab{\e_{K}}^{\rm exp}=2.228(11)\times 10^{-3}.
\end{equation}
On the other hand, there are larger uncertainties in the theoretical predictions of the SM \cite{Jang:2017ieg}.
Using exclusive $V_{cb}$ determined by lattice QCD, the SM prediction is 
 \begin{equation}
\ab{\e_{K}}^{\rm SM,ex}=1.58 (16) \times 10^{-3}.
\end{equation} 
Using inclusive $V_{cb}$ determined by QCD sum-rule, 
\begin{equation}
\ab{\e_{K}}^{\rm SM,in}=2.05 (18) \times 10^{-3}.
\end{equation}
Although $\ab{\e_{K}}^{\rm SM,in}$ is consistent with the experiment, {$\ab{\e_{K}}^{\rm SM,ex}$  has a deviation around $4\s$ 
level.}

\begin{figure}[!t]
\begin{center}  
   \includegraphics[width=75mm]{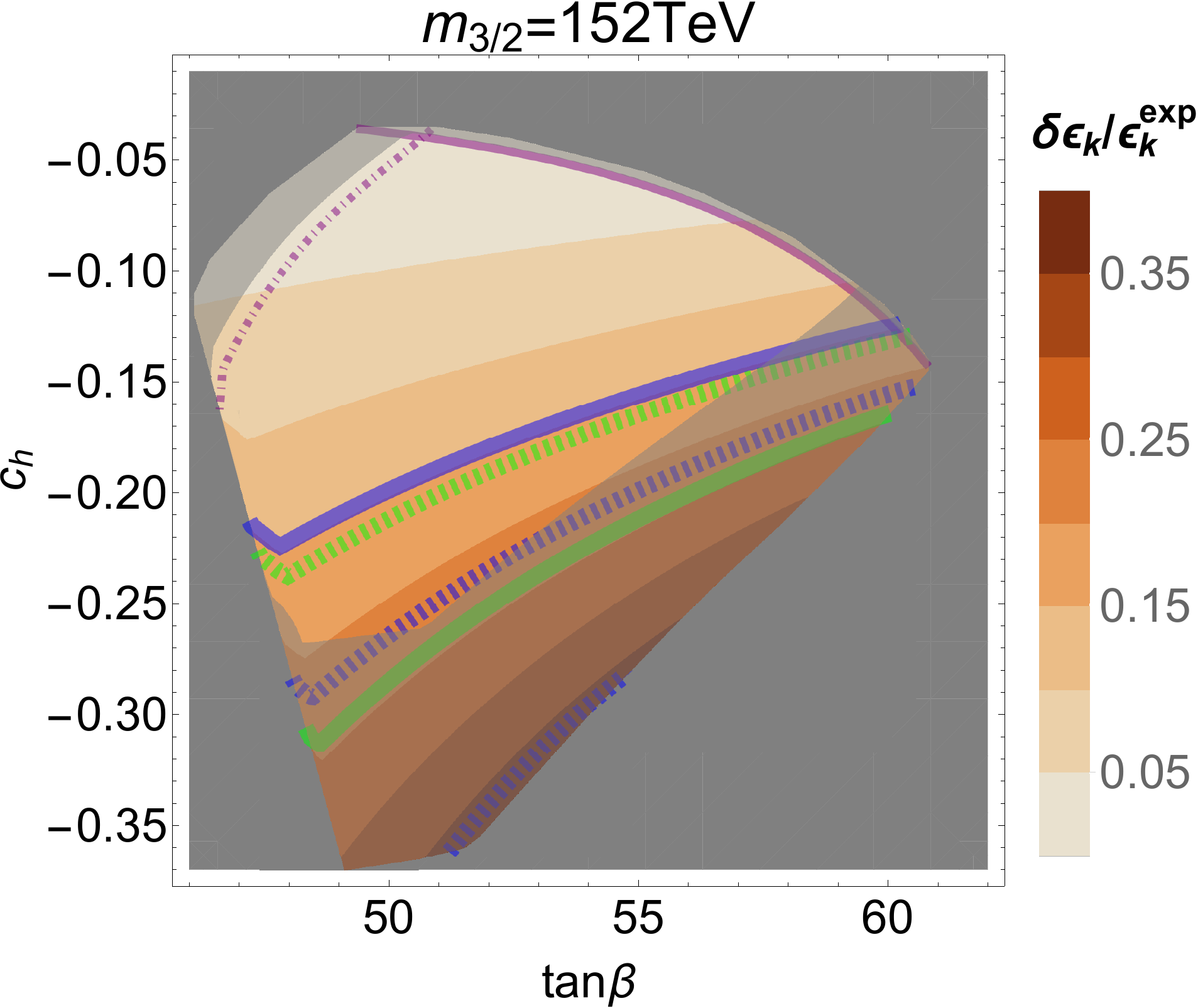}
   \includegraphics[width=75mm]{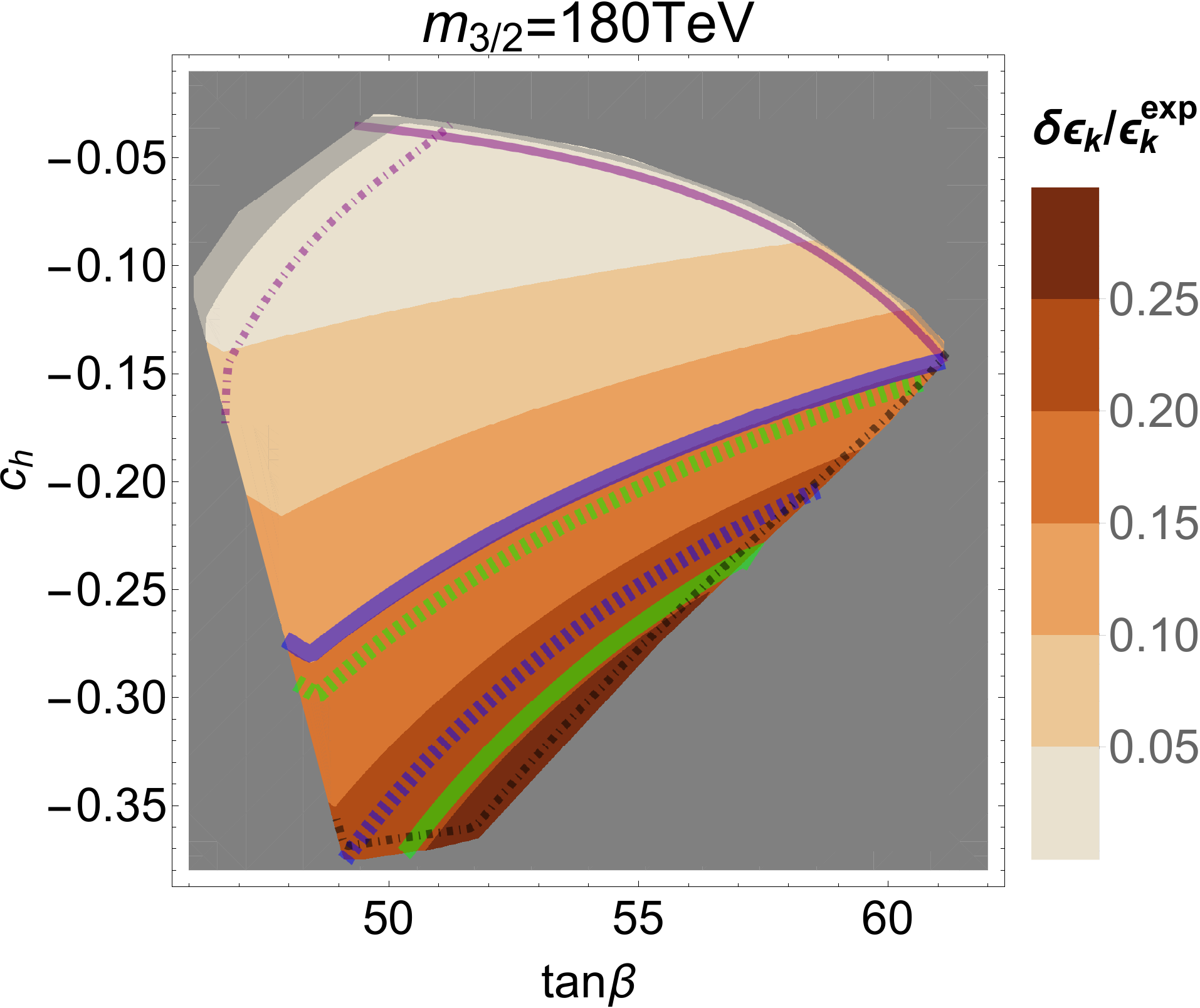}
\end{center}
\caption{ Contours of the ratio of SUSY contribution to the experimental value of $\e_k$ for $m_{3/2}=152\TEV$(left) and $m_{3/2}=180\TEV$(right). The gray shaded regions denote the LHC bounds. On the purple dash-dotted (solid) line, the mass of the right (left)-handed selectron degenerates with the wino mass. 
On the black dash-dotted line in the right panel, the mass of the lightest squark degenerates with the wino mass.
Below (Above) the blue (green) solid and dashed lines, the result is consistent with $\e_k^{\rm exp}$ at $2\s$ and $1\s$ level, using the exclusive (inclusive) $V_{cb}$.
}
\label{fig:4}
\end{figure}

In Fig.\ref{fig:4} the ratio of $\d \e_K$ to $\e_{K}^{\rm exp}$ is represented, where $\d \e_K$ is the SUSY contribution to $\e_{K}$. 
In the shaded-gray region, the slepton mass is $\lesssim 440\GEV$ or the squark mass is $\lesssim 1400\GEV$ representing the LHC bound for the direct slepton production followed by its $R$-parity-violating leptonic decay~(c.f. Ref. \cite{CMS:2017mkt}) or direct squark production~\cite{Aaboud:2017vwy, Sirunyan:2017cwe}.\footnote{We note that this bound is set with the assumption that all the masses of the first and second generation sleptons (squarks) are degenerate. This may be an overestimation because the heavier sleptons (squarks) reduce the sparticle production rates compared with those in the references.} 
Now one finds that the SUSY contributions in all the region are either consistent with the experiment regarding ${\e_{K}}^{\rm SM,ex}$ or ${\e_{K}}^{\rm SM,in}$. 
Interestingly, the deviation, $\d \e_K$, can be of $\O(10)\%$ of the experimental value. 
 The discrepancy between $\e_{K}^{\rm SM,ex}$ and $\e_{K}^{\rm exp}$ can be filled for large $|c_h|$ and $\tan\b$. 
In this case, the lightest squark and wino has almost degenerate masses, which is within the reach of the 33 TeV collider. 
We have also checked that there are no significant deviations from the SM predictions in the $D \AND B$-meson system  as well as $\D M_K$ and $\e'_K/\e_K$ \cite{Kagan:1999iq}.\footnote{We have also checked that $\D M_K, \e_K$ and $\e'_K/\e_K$ do not significantly exceed the SM prediction with $\m_0<0$.} 
\\

So far, we have shown that the Higgs-Anomaly mediation solves the SUSY FCNC problem. This is 
quite non-trivial, 
because the setup has sufficiently heavy stops favored by the measured Higgs boson mass and various light sparticles which can affect low energy phenomena. Some phenomena will be discussed in the next section.

\section{Other Predictions}
\lac{pred}
As mentioned, we have modified the matching scale in $\tt SuSpect$ taking account of the sparticle mass splitting. 
By correctly setting the matching scales, $y_b$ in the MSSM is reduced from the previous analysis in Refs.\cite{Yin:2016shg, Yanagida:2016kag} for the fixed $\tan\beta$ and the stop mass scale. 
As a result, viable parameter regions slightly move towards larger $\tan\b$. 
Due to this modification, we will see that the region explaining the muon $g-2$ anomaly becomes larger due to the $\tan\b$ enhancement, while the region consistent with the Yukawa coupling unification moves towards smaller $\ab{c_h}$. 
The muon $g-2$ anomaly and the Yukawa coupling unifications will be obtained for $\m_0>0$, but the threshold corrections relevant for them change their signs for $\m_0<0$ [see \Eqs{gm2} and \eq{th}]. The relative sign between gluino and bino masses due to anomaly mediation will also be relevant in explaining the muon $g-2$ anomaly and Yukawa unifications simultaneously.

\paragraph{Muon $g-2$}

  \begin{figure}[!t]
\begin{center}  
   \includegraphics[width=80mm]{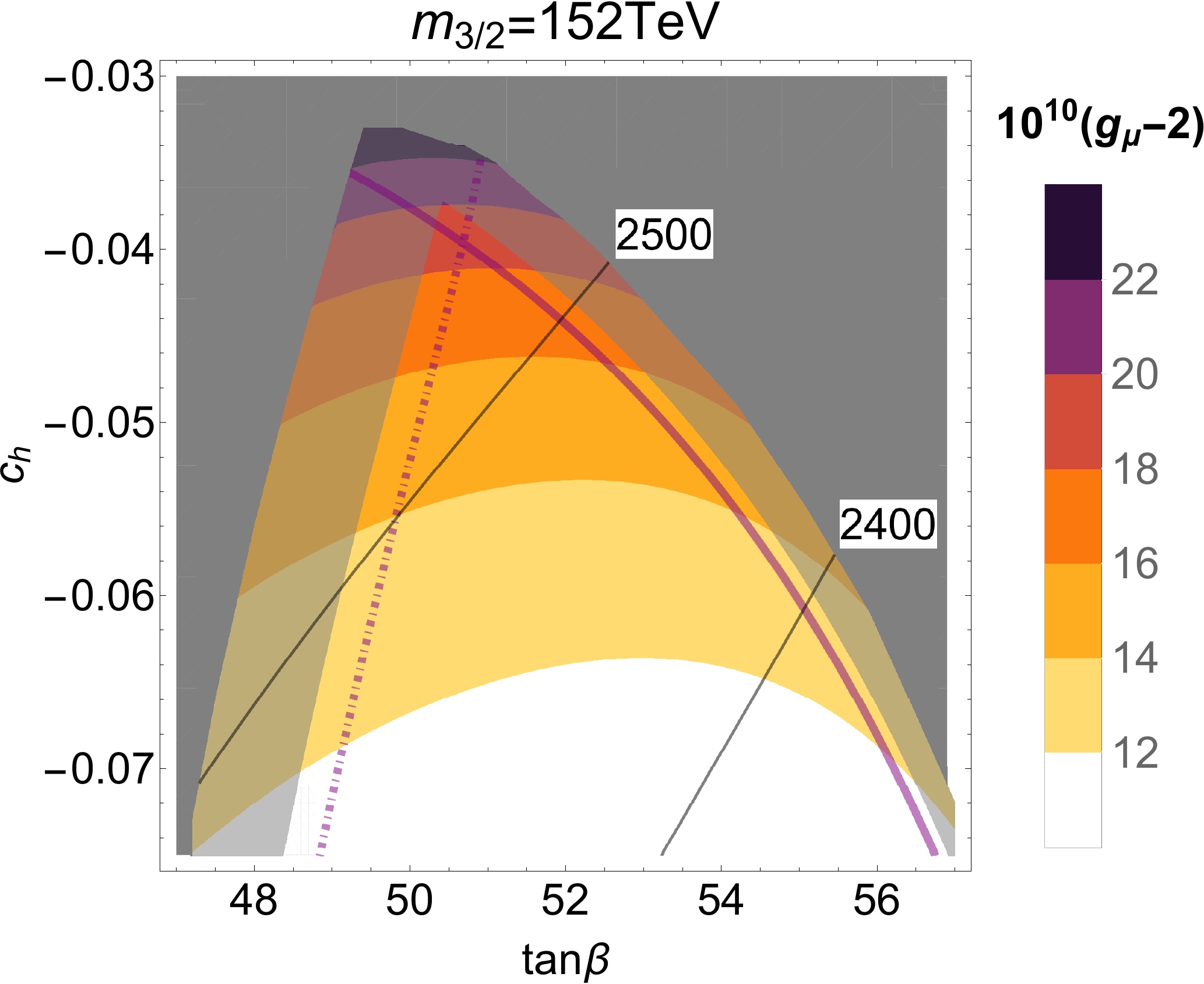}
   \includegraphics[width=80mm]{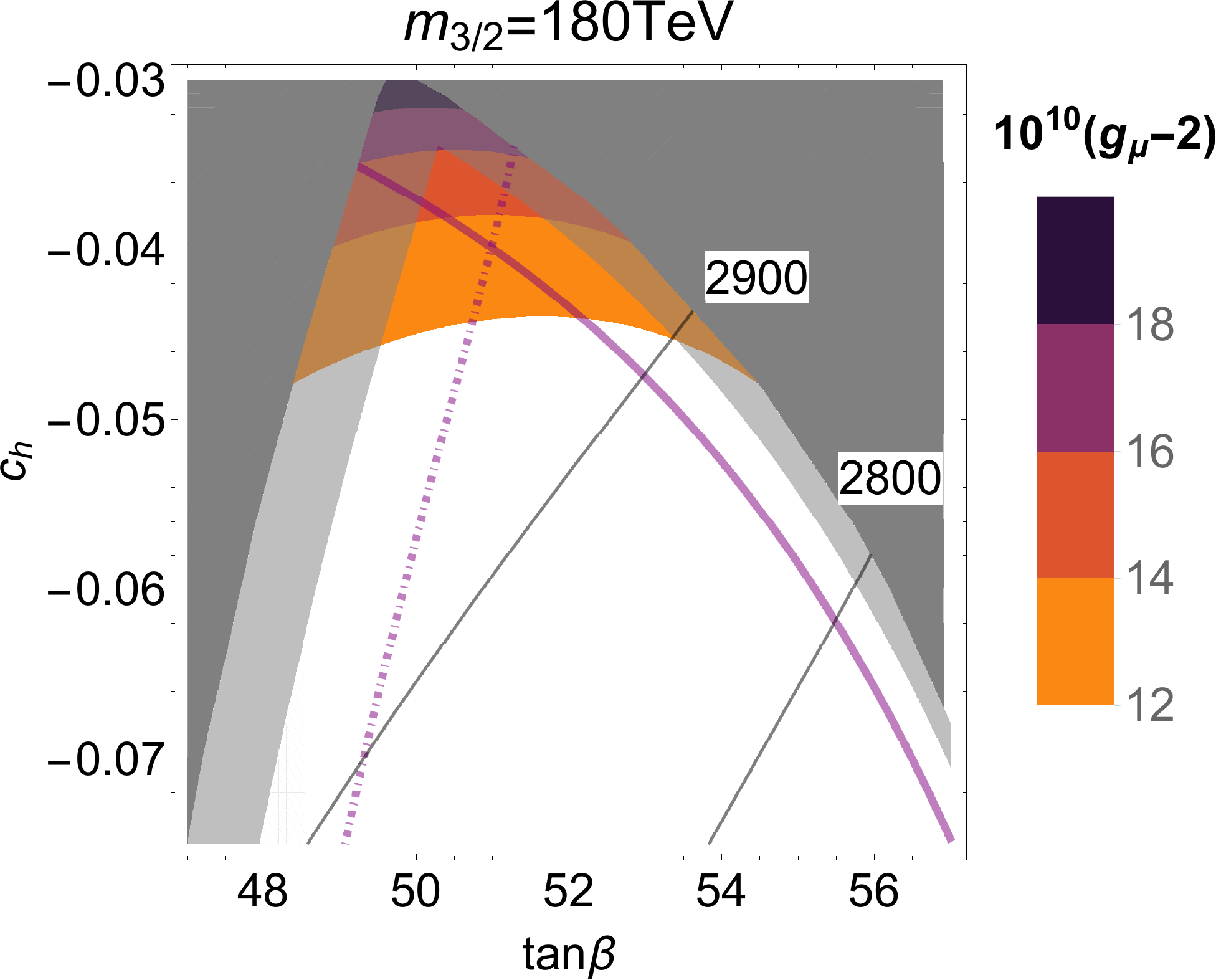}
\end{center}
\caption{ Contours of the SUSY contribution to the muon $g-2$ for $m_{3/2}=152\TEV$ (left) and $m_{3/2}=180\TEV$ (right). The geometric mean of the first two generation squark masses [GeV] is shown in black solid line. 
The gluino masses are $\approx 3.2\TEV$ (left) and $\approx 3.7\TEV$ (right).} 
\label{fig:muong2}
\end{figure}

The Higgs mediation effect is essential in the lepton sector to solve the tachyonic-slepton problem\cite{Yin:2016shg, Yanagida:2016kag}. 
One has light sleptons whose masses are small due to the negative contribution from anomaly mediation especially for $c_h=-\O(0.01)$. 
This leads to the attractive feature: the explanation of the muon $g-2$ anomaly.\footnote{
SUSY models beyond the MSSM explaining the muon $g-2$ anomaly are shown in Refs.~\cite{Endo:2011mc, Endo:2011xq, Endo:2011gy, Moroi:2011aa, Nakayama:2012zc, Sato:2012bf, Shimizu:2015ara, Yin:2016pkz, Higaki:2016yeh, Fukuyama:2016mqb} 
} Since $\m \AND \tan\b$ are large which are required for a correct EWSB, the contribution to the muon $g-2$ from the bino-smuon loop is important\footnote{It is checked that although we have a quite large $\m \tan\b$ and light smuons, the EWSB vacuum is enough long-lived in all the viable region throughout the thermal history of the universe (c.f. Ref.\cite{Endo:2013lva}).}.
This contribution can be approximated as~\cite{Lopez:1993vi, Chattopadhyay:1995ae, Moroi:1995yh}
\begin{eqnarray}
(\alpha_\mu)_{\rm SUSY} \simeq \left(
\frac{1 - \delta_{\rm QED}}{1 + \Delta_\mu }
\right) 
{g_Y^2 \over 16\pi^2}{ m_\mu^2  \mu \tan\beta \, M_1 \over m_{\tilde{\mu}_L}^2 m_{\tilde{\mu}_R}^2}
\,f_N\left( 
\frac{m_{\tilde{\mu}_L}^2}{M_1^2},
\frac{m_{\tilde{\mu}_R}^2}{M_1^2}
\right) , \label{eq:gm2}
\end{eqnarray}
where $m_{\mu}$ is the muon mass; 
$m_{\tilde{\mu}_L}$ $(m_{\tilde{\mu}_R})$ is the mass of the left-handed  (right-handed) smuon;
$f_N(x,y)$ is a loop function of $\mathcal{O}(0.1)$; $\Delta_\mu$ and $\delta_{\rm QED}$  are two-loop corrections given in~\cite{Marchetti:2008hw,Degrassi:1998es} which are of $\mathcal{O}(0.1)$.
On the other hand the anomaly is represented by
\begin{eqnarray}
a_\mu^{\rm EXP} - a_\mu^{\rm SM} = (26.1 \pm 8.0) \times 10^{-10},
\end{eqnarray}
where we have quoted~Ref.\cite{Hagiwara:2011af} for a SM prediction \cite{Davier:2010nc, Hagiwara:2011af }, $a_\m^{\rm SM}$, 
while $a_\mu^{\rm EXP}$ is the experimental value~\cite{Bennett:2006fi,Roberts:2010cj}. 

In Fig.\ref{fig:muong2}, one finds that the muon $g-2$ anomaly can be explained at $1\s$ ($2\s$) level for $m_{3/2}\lesssim 150\,\TEV$ $(190\, \TEV)$ when the lightest sparticle (LSP) in the MSSM is wino-like. The region within $1\s$ ($2\s$) level can (could) be within the reach of the high-luminosity LHC, since the masses of the light squarks are almost degenerate. This region can be also fully tested from the search of disappearing charged track for wino-like LSP \cite{Fukuda:2017jmk} (see also Ref. \cite{Asai:2007sw, Asai:2008sk})

When $R$-parity is violated or there is a lighter sparticle in a different sector, selectron can be lighter than the wino.\footnote{Then the dark matter could be QCD axion/axino, inflaton/inflatino \cite{Kofman:1994rk, Kofman:1997yn, Lerner:2009xg, Okada:2010jd, Khoze:2013uia, Bastero-Gil:2015lga, Daido:2017wwb, Nakayama:2010kt, Daido:2017tbr} or a light singlet predicted in the pseudo Nambu-Goldstone boson hypothesis for sfermions (c.f. Ref.\cite{Harigaya:2015iva,Yanagida:2016kag}).} In this case, the muon $g-2$ region can be enlarged with heavier gravitino $m_{3/2}\lesssim 160\,\TEV$ $(200\, \TEV)$ for the explanation at $1\s$ (2$\s$) level. Such selectrons could be sufficiently long-lived, and within the reach of the LHC \cite{llstau,llstau2,llstau3}.

\begin{figure}[!t]
\begin{center}  
   \includegraphics[width=80mm]{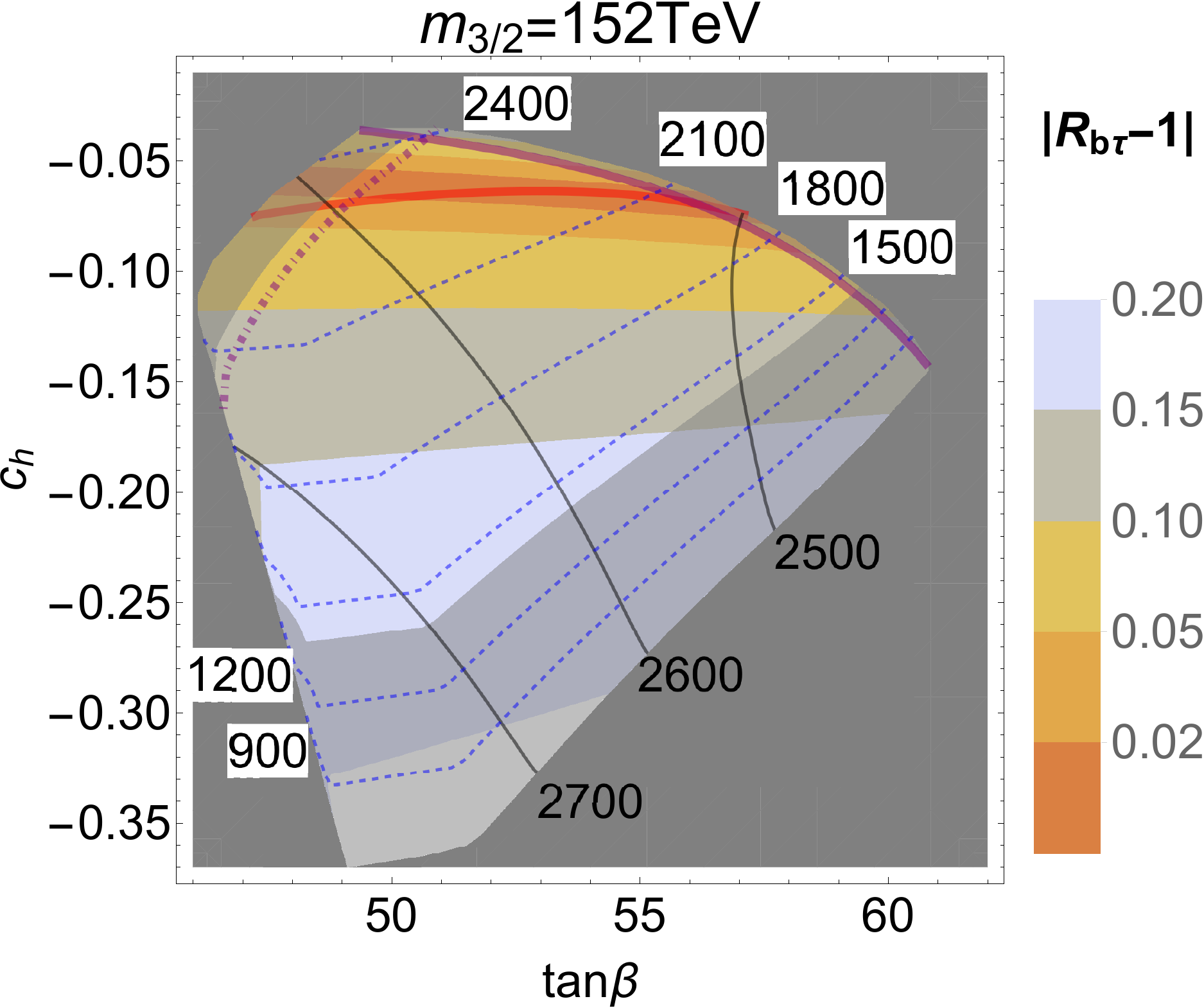}
   \includegraphics[width=80mm]{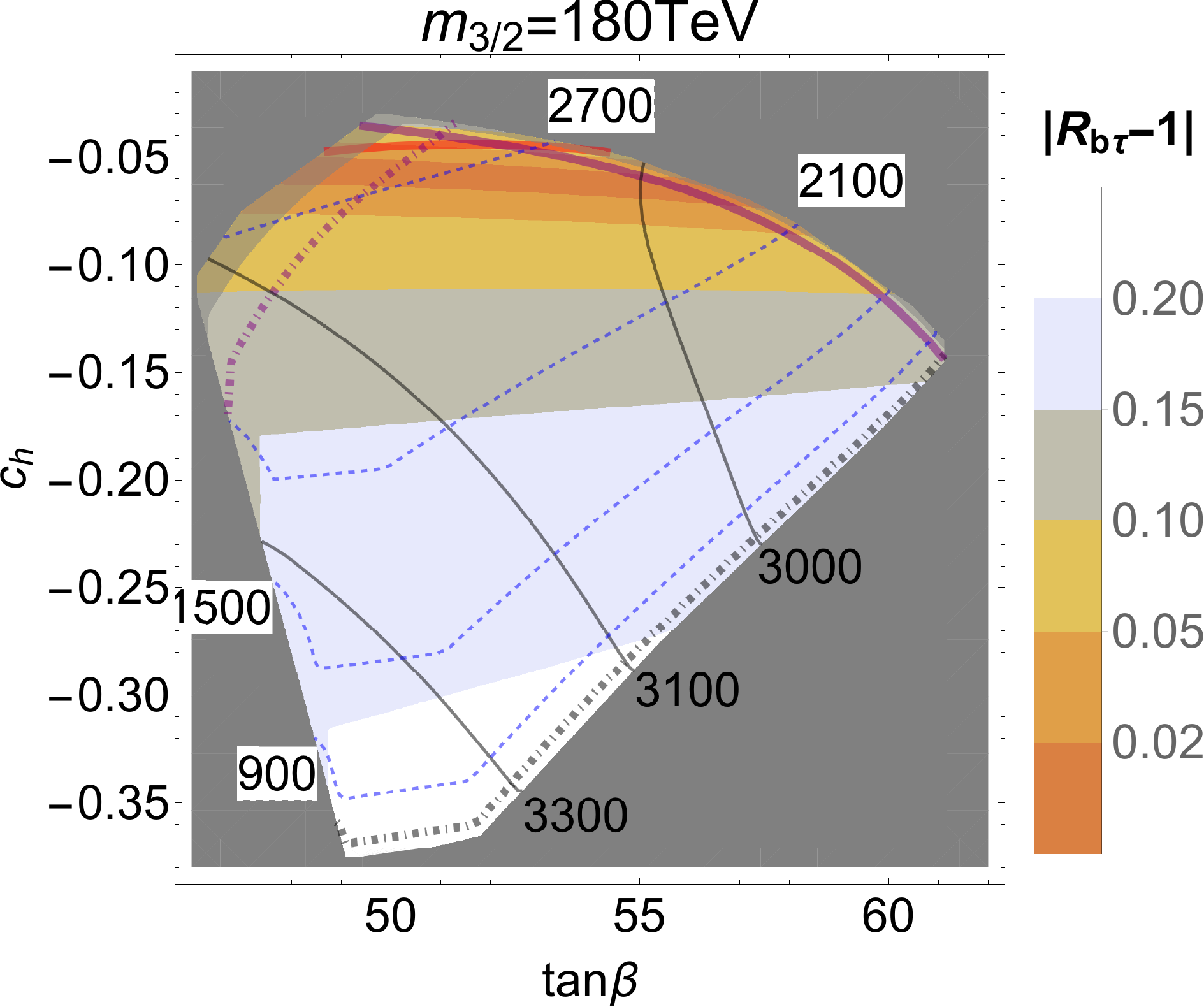}
\end{center}
\caption{ {Contours of $\ab{R_{b\tau}-1}$ for $m_{3/2}=152\TEV$ (left) and $180\TEV$ (right).} The black solid (dashed) contours represent the mass of the lightest (heaviest) squark in the first two generation in units of GeV. Above the red solid line the muon $g-2$ anomaly is explained within the $2\s$ level.}
\label{fig:3}
\end{figure}

\begin{figure}[!t]
\begin{center}  
\includegraphics[width=80mm]{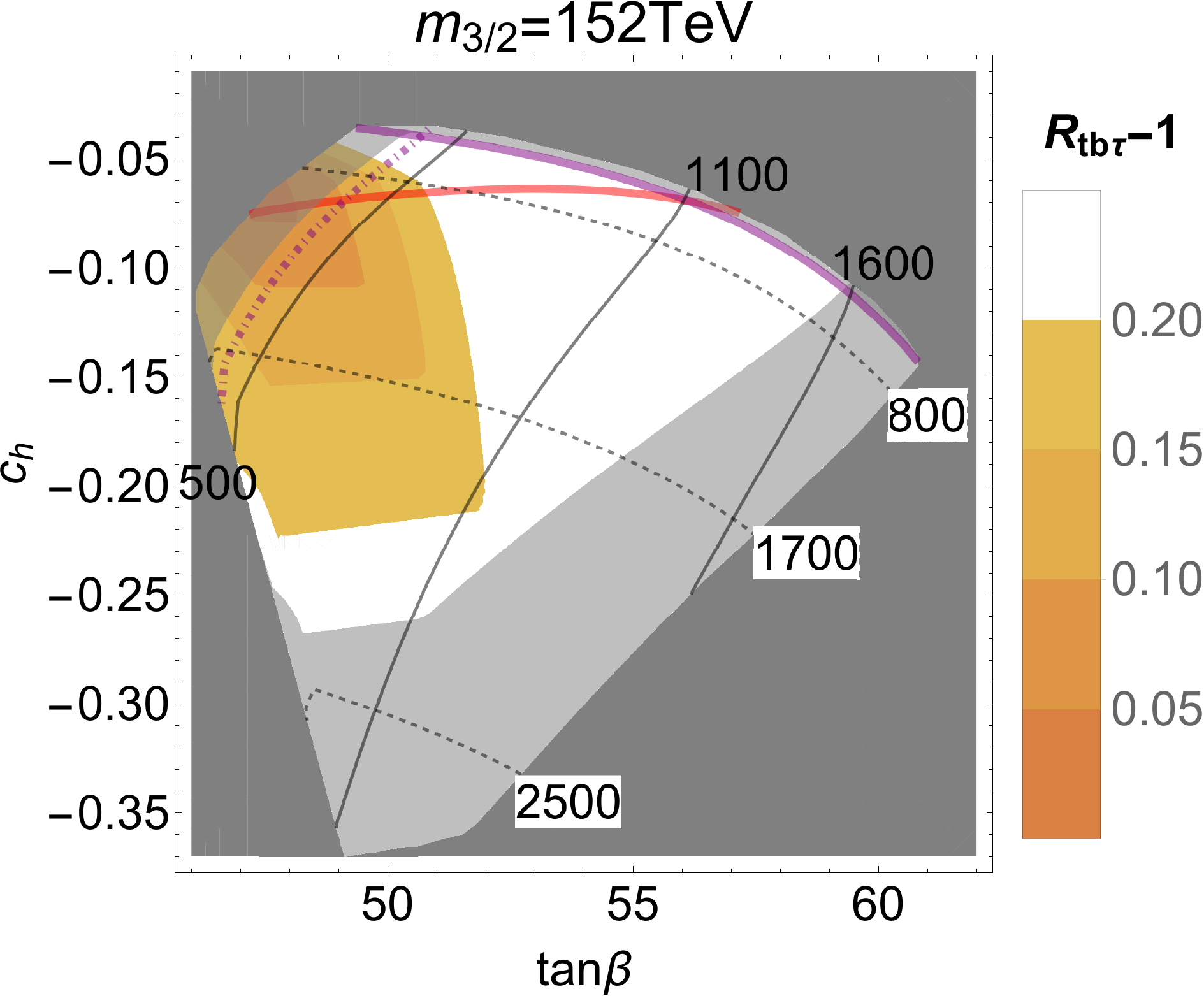}
   \includegraphics[width=80mm]{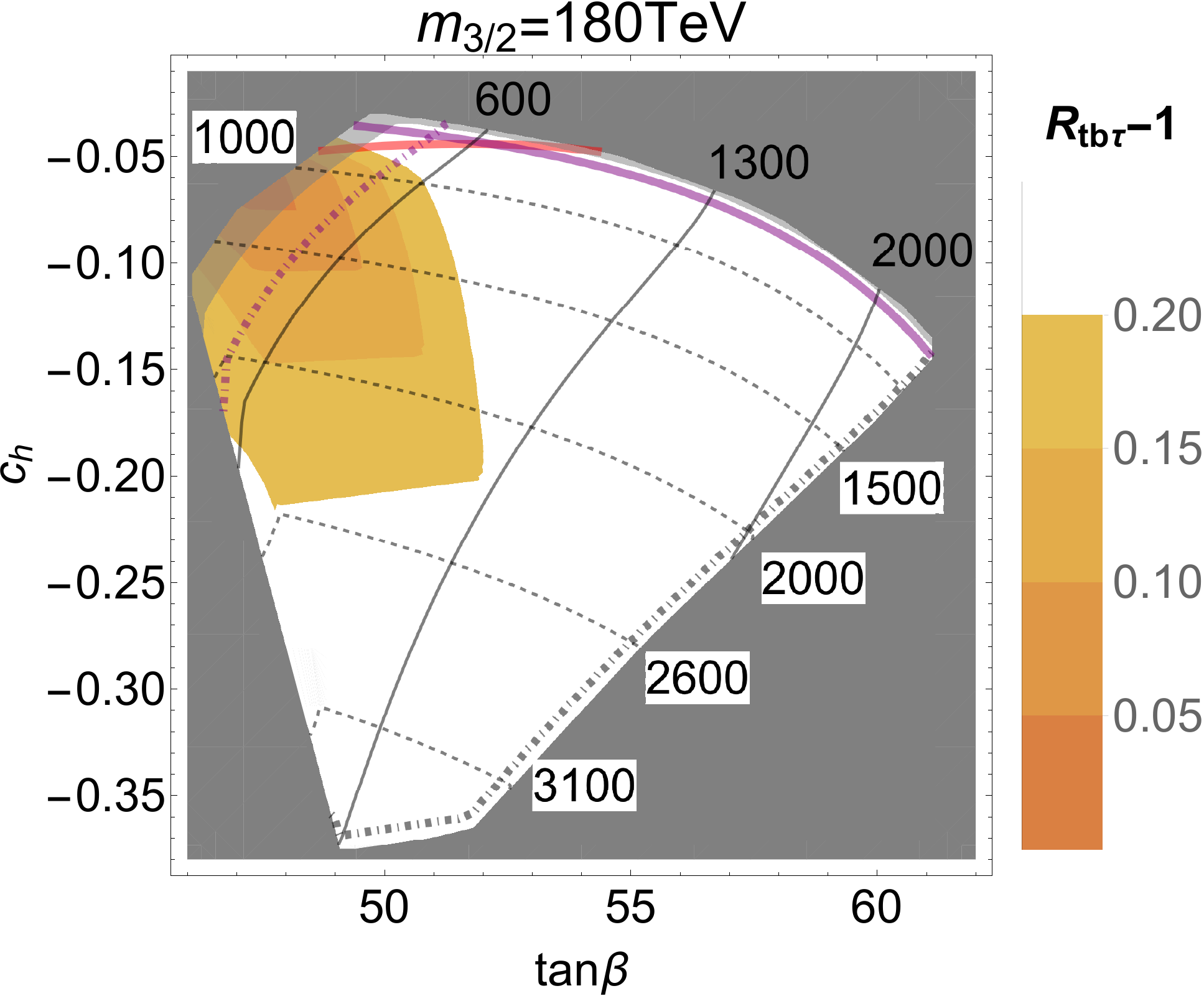}
\end{center}
\caption{ Contours of $R_{tb\tau}-1$ for $m_{3/2}=152\TEV$ (left) and $180\TEV$ (right). 
The black solid (dashed) line represents the mass of the right (left)-handed selectron in units of GeV. }
\label{fig:SO10}
\end{figure}

\paragraph{Yukawa Unification} 
Since sizable $y_b \AND y_\t$ are required for the successful EWSB, one can have Yukawa coupling unifications at $M_{\rm inp}$.\footnote{We do not require the first two generation Yukawa couplings to unify because of corrections from higher dimensional operators of the order of $M_{\rm inp}/M_{\rm pl}\sim 10^{-2}$. 
$M_{\rm inp}/M_{\rm pl}$ is also the order of the one-loop threshold corrections to the gauge and third generation Yukawa couplings at $M_{\rm inp}$. }
The precisions of the bottom-tau and top-bottom-tau Yukawa unifications are defined by
\begin{equation}
R_{b\t}={y_b \o y_\t}
\end{equation}
and 
\begin{equation}
R_{tb\tau}={\max\(y_t,y_b,y_\t\) \o \min\(y_t,y_b,y_\t\) }
\end{equation}
 at $M_{\rm inp}$, respectively. The contours of $R_{b\t}$ and $R_{tb\tau}$ are, respectively, shown in Fig.\ref{fig:3} and Fig.\ref{fig:SO10}.\footnote{We note that in an $\SO(10)$ GUT, a right-handed neutrino as well as the matters in a generation is embedded in a $\bf 16$ multiplet. This implies the neutrino Yukawa coupling in $y_\n^{i,j} H_u L_i \n_j^R$ would affect the RG-running of the $y_t$ and $y_\tau$. Thus from the viewpoint of $\SO(10)$ GUT, $y_t$ and $y_\tau$ could be raised slightly at the GUT scale and the region for the top-bottom-tau Yukawa unification could move towards larger $\tan\b$ and smaller $\ab{c_h}$. This effect may enlarge the viable region for the unification. }
The masses of relevant sfermions are also shown. 
Almost degenerated squarks in the region consistent with the Yukawa unification can be searched at the LHC as well as the future collider experiments (see \Sec{lsq}).
The red solid line denotes the $2\s$ boundary of the muon $g-2$. 
{One can find that there are regions both favored by the muon $g-2$ and the Yukawa coupling unifications.  These regions are within the reach of the LHC. 
}Note that the regions consistent with the Yukawa coupling unifications do not depend much on the size of $m_{3/2}$ and can be found even for $m_{3/2} \simeq \O(10^3)\TEV$.

\begin{table*}[!t]
\caption{\small Mass Spectra for some model points.
}
\label{tab:1}
\begin{center}
\begin{tabular}{|c||c|c|c|c|}
\hline
Parameters & Point {\bf I} & Point {\bf II}  & Point {\bf III} & Point {\bf IIII} \\
\hline
$m_{3/2} $ (TeV) & 150  & 151& 180 & 250  \\
$c_h$  & $-0.041$  & $-0.07$  & $-0.35$& $-0.053$\\
$\tan\beta$  & 50.8  & 48.9  & 51.0 & 52.0 \\
LSP  & $\chi_1^0$  & ${\tl{e}}_R$  & $\chi_1^0$ & $\chi_1^0$ \\
\hline
Particles & Mass (GeV) & Mass (GeV)& Mass (GeV) & Mass (GeV)  \\
\hline
$\tilde{g}$ (TeV)& 3.12 & 3.15 & 3.77 &  4.98  \\
$\tilde{\chi}_{1,2}^0$ & 475, 1380 & 482, 1390 & 577, 1680 & 785, 2310  \\
$\tilde{u}_{6,5}$ (TeV) & 13.2, 12.8 & 16.4, 16.2 & 41.8, 39.5 & 24.2 23.7\\
$\tilde{d}_{6,5}$ (TeV) & 14.0, 13.5 & 16.8, 16.5 & 41.1, 38.7 & 25.6, 24.8 \\
$\tilde{u}_{4,3}$  &2580, 2560& 2620, 2600 & 3360, 3240 & 4140, 4110\\
$\tilde{d}_{4,3}$ &2580, 2560& 2620, 2600 & 3360, 3240 &4140, 4110\\
$\tilde{u}_{2,1}$ &2360, 2360& 2300, 2300 & 1050, 955 & 3670, 3660\\
$\tilde{d}_{2,1}$ &2470, 2470& 2370, 2360 & 788, 783 & 3880, 3870\\
$\tilde{e}_{L, R}$ &497, 496& 1030, 467& 1570, 3230& 1130, 968\\
$\tilde{\mu}_{2,1}$ &609, 557& 1090,  665& 3400, 2170& 1230, 1190\\
$\tilde{\tau}_{2,1}$ (TeV) & 12.2, 8.6& 15.1, 10.7& 40.8, 28.9&  23.4, 16.5 \\
$H^\pm$\,(TeV) & 10.5 & 9.7 & 16.0 & 19.1 \\
$h_{\rm SM\mathchar`-like}$ (${\tt HD}$) & 120.3 &  121.4  & 124.4 & 122.5\\
$h_{\rm SM\mathchar`-like}$ (${\tt FH}$) & 124.7 &  125.9  &  133.9 & 127.6 \\
$10^{9}\d \a_\mu $& 1.85 &  1.21  & 0.22 & 0.53\\
$\d \e_K/\e_K^{\rm SM}$& 1.2~\% &  2.7~\%  & 24~\% & 0.8~\% \\
$R_{b\t}$& 1.08 &  0.97 & 0.79 &1.00 \\
$R_{tb\t}$& 1.37 &  1.09  & 1.27 &1.29 \\
\hline
$\tl{\chi}^{\pm}_2$ (TeV) & 25.7 & 34.0  & 91.8 &  49.0\\
\hline
\end{tabular}
\end{center}
\end{table*}

\vspace{12pt}
Finally, some data points are shown in Table \ref{tab:1}.
On the point {\bf II}, 
we consider that the LSP of the MSSM is the selectron with assuming $R$-parity violation to survive the experimental/cosmological bound while the LSP is wino-like neutralino on the other points.
On the point {\bf I} ({\bf II}), the muon $g-2$ anomaly is explained at 1$\sigma$ (2$\sigma$) level.
On all the points, the stop-like squarks are heavy as $\mathcal{O}(10)$\,TeV, which are expected to be consistent with the observed Higgs boson mass of 125\,GeV.

\section{Discussion and Conclusions}

We have shown that the SUSY flavor problem is solved in a simple 
setup called Higgs-Anomaly Mediation.
This simple setup has splitting mass spectra of sfermions: the masses of the third generation sfermions are of $\O(10)\TEV$, and those of the first and second generation sfermions are of $\O(1)\TEV$ for the gravitino mass of $\O(100)\TEV$. Consequently, the scenario is not only consistent with the measured Higgs boson mass but also provides many interesting features in ground-based experiments. Moreover, the gravitino problem is solved since the gravitino decays before the big bang nucleosynthesis.

Performing the precise analysis in the mass eigenbasis, 
we found that the mass-splitting in the spectra even suppresses the flavor-changing processes:
{the SUSY contribution to the $K$-meson mixing is less than $\mathcal{O}(10)\%$ of the estimation made by a na\"{i}ve mass insertion approximation assuming no cancelation between diagrams.} Therefore, our setup is flavor-safe in all viable regions.

Surprisingly, the Higgs-Anomaly mediation can explain the muon $g-2$ anomaly and the top-botton-tau/bottom-tau Yukawa coupling unification simultaneously. We revisited these phenomena with precise analyses taking account of the off-diagonal elements of the mass matrices and the renormalization scales of the threshold corrections. 
The masses of the first two generation squarks are below $\sim 2.5 \TEV$ $(3\TEV)$ in the regions where the muon $g-2$ anomaly is explained at $1\s$ (2$\s$) level. 
These light squarks could be fully tested in the (high-luminosity) LHC.

\section*{Acknowledgments}
We thank S. Matsumoto for fruitful discussion on pure wino dark matter and its detectability, and S. Shirai for the useful discussion about the two-loop threshold correction effect on the Higgs boson mass. 
We are also grateful to H. Murayama for pointing out that the bulk loop correction on the flavor-violating parameters in the extra dimension scenario. 
W.Y would like to thank Kavli IPMU for the kind hospitality during the final stage of this work.
This work is supported by JSPS KAKENHI Grant Numbers JP26104009 (T.T.Y), JP26287039 (T.T.Y.), JP16H02176 (T.T.Y), JP15H05889 (N.Y.), JP15K21733 (N.Y.), JP17H05396 (N.Y.), JP17H02875 (N.Y.), and by World Premier International Research Center Initiative (WPI Initiative), MEXT, Japan (T.T.Y.).

\section*{Appendix A: Analytic Formulas $K$ meson mixing}

We show approximated analytic formulae for dominant corrections to the kaon mass mixing parameter.(c.f. Ref.\cite{Gabbiani:1996hi, Goto})
\begin{equation}
{\d M_{12}^{K}}\simeq { 1 \over 3} {B_1^K} f_K^2 {m_K} \eta_{VLL} \(\d_{\tl{g}}+\d_{\tl{W}^{\pm}}+\d_{{\tl{g}\tl{W}^0}}\)
\end{equation}
where $\d_{\tl{g}}$, $\d_{\tl{W}^{\pm}}$ and $\d_{{\tl{g}\tl{W}^0}}$ are the gluino, charged wino, and gluino-wino contributions, respectively. They are given at the renormalization scale $M_3$ as
\begin{equation}
\d_{\tl{g}}\simeq -{\alpha_3^2\o 108 \tr[{\bf m}^{2,\rm eff}_{\tl{Q}}]}  \left(24 x_{\tl{g}} f\left( x_{\tl{g}} \right)+66 {\tl{f}}\left(x_{\tl{g}} \right)\right) \( \(\d^{d,\rm eff}_{QLL}\)_{1,2} \)^2,
\end{equation}
\begin{equation}
\d_{\tl{W}^{\pm}}\simeq-{\alpha_2^2\o 108 \tr[{\bf m}^{2,\rm eff}_{\tl{Q}}]}\left(27 {\tl{f}}\left(x_{\tl{W}^{\pm}} \right)\right) \( \(\d^{d,\rm eff}_{QLL}\)_{1,2} \)^2,
\end{equation}
\begin{equation}
{\d_{\tl{g}\tl{W}^0}}\simeq
\laq{mixed}
-{\alpha_2 \a_3 \o 108 \tr[{\bf m}^{2,\rm eff}_{\tl{Q}}]} \left(72 {M_2\o M_3} h\left(1/x_{\tl{g}},x_{\tl{g}\tl{W}^{0}}\right)+36\tl{h} \left(1/x_{\tl{g}},x_{\tl{g}\tl{W}^{0}} \right)\right)\( \(\d^{d,\rm eff}_{QLL}\)_{1,2} \)^2,
\end{equation}
where $x_{\tl{g}}\equiv {2M_3^2 \o \tr[{\bf m}^{2,\rm eff}_{\tl{Q}}]}$ and $x_{\tl{g}\tl{W}^0}\equiv {M_2^2 \o M_3^2}$; $B_1^K\simeq 0.76$ (the bag parameter), $f_K\simeq 0.16\GEV$ and $m_K\simeq 0.50\GEV$ are found in \cite{Aoki:2016frl}. 
{ To match the Wilson coefficient at the scale where $B_1^K$, $f_K$ and $m_K$ are defined, the wave function renormalization at one-loop level is included:}
\begin{equation}
\eta_{VLL}= \left(\frac{{\alpha_3({M_{3}})}}{{\alpha_3({m_{t}})}}\right)^{6/21} \left(\frac{{\alpha_3({m_t})}}{{\alpha_3({m_b})}}\right)^{6/23}\left(\frac{{\alpha_3({m_b})}}{{\alpha_3({\rm 3~GeV})}}\right)^{6/25}.
\end{equation}
Here, the $\SU(3)$ coupling constants at the top mass scale, bottom mass scale and $3\GEV$ are $\alpha_3({m_{t}})\simeq 0.11, \alpha_3({m_{b}})\simeq 0.21$ and $\alpha_3({3\GEV})\simeq 0.25$, respectively.
The loop functions are given by,
\begin{equation}
f(x)=\frac{x^3-9 x^2-9 x+6 (3 x+1) \log (x)+17}{6 (x-1)^5},\end{equation}\begin{equation} \tl{f}(x)=\frac{-x^3-9 x^2+9 x+6 (x+1) x \log (x)+1}{3 (x-1)^5},\end{equation}\begin{equation}
h(x,y)=\left.-x^3 {\partial^2 g_1(x,z,y) \o \partial z^2}\right|_{z=x},\end{equation}\begin{equation} 
\tl{h}(x,y)=\left. -x^3{\partial^2 g_2(x,z,y) \o \partial z^2}\right|_{z=x},
\end{equation}
where
\begin{equation} g_1(x,z,y)=\frac{x \log (x)}{(x-1) (x-y) (x-z)}+\frac{y \log (y)}{(y-1) (y-x) (y-z)}+\frac{z \log (z)}{(z-1) (z-x) (z-y)}\end{equation} 
and
\begin{equation} g_2(x,z,y)=\frac{x^2 \log (x)}{(x-1) (x-y) (x-z)}+\frac{y^2 \log (y)}{(y-1) (y-x) (y-z)}+\frac{z^2 \log (z)}{(z-1) (z-x) (z-y)}.\end{equation} 
We find $\d_{\tl{g}}$ is comparable with the other two due to the cancellation between $f(x_{\tl{g}}) \AND \tl{f}(x_{\tl{g}})$. The cancellation does not occur in $\d_{\tl{W}^{\pm}}$ and $\d_{\tl{g}\tl{W}^0}$.

$\d M_{12}^K$ is related to the contribution to the mass difference of $K$-meson as
\beq
\d \D M_K=  2\Re[\d M_{12}^K],
\eeq
and that to the CP phase as 
\beq
\d \e_K \simeq {\exp{(i\f_\e)}\sin(\f_\e)\over \D M_{K}^{\rm exp}}{\Im[\d M_{12}^K]}.
\eeq
Here,  
 $\D M_{K}^{\rm exp}\simeq 3.5 \times 10^{-15} \GEV $ and $\f_\e \simeq 44^\circ$.

\providecommand{\href}[2]{#2}\begingroup\raggedright\endgroup


\begin{thebibliography}{10}

\bibitem{Inoue:1991rk} 
  K.~Inoue, M.~Kawasaki, M.~Yamaguchi and T.~Yanagida,
  Phys.\ Rev.\ D {\bf 45}, 328 (1992).


 \bibitem{Randall:1998uk} 
  L.~Randall and R.~Sundrum,
  Nucl.\ Phys.\ B {\bf 557}, 79 (1999)
  [hep-th/9810155].

  
  
\bibitem{Giudice:1998xp} 
  G.~F.~Giudice, M.~A.~Luty, H.~Murayama and R.~Rattazzi,
  JHEP {\bf 9812}, 027 (1998)
  [hep-ph/9810442].
   
   

  
\bibitem{Yin:2016shg} 
 W.~Yin and N.~Yokozaki,
  Phys.\ Lett.\ B {\bf 762}, 72 (2016)
  [arXiv:1607.05705 [hep-ph]].


\bibitem{Yanagida:2016kag} 
  T.~T.~Yanagida, W.~Yin and N.~Yokozaki,
  JHEP {\bf 1609}, 086 (2016)
  [arXiv:1608.06618 [hep-ph]].
  
    \bibitem{Ibe:2006de} 
  M.~Ibe, T.~Moroi and T.~T.~Yanagida,
  Phys.\ Lett.\ B {\bf 644}, 355 (2007)
  [hep-ph/0610277].

  \bibitem{Ibe:2011aa} 
  M.~Ibe and T.~T.~Yanagida,
  Phys.\ Lett.\ B {\bf 709}, 374 (2012)
  [arXiv:1112.2462 [hep-ph]].
  
  
  
    \bibitem{ArkaniHamed:2012gw} 
  N.~Arkani-Hamed, A.~Gupta, D.~E.~Kaplan, N.~Weiner and T.~Zorawski,
  arXiv:1212.6971 [hep-ph].
  


  
  
 \bibitem{Okada:1990vk}
Y.~Okada, M.~Yamaguchi and T.~Yanagida,
  Prog.\ Theor.\ Phys.\  {\bf 85}, 1 (1991).
  
  
\bibitem{Ellis:1990nz} 
  J.~R.~Ellis, G.~Ridolfi and F.~Zwirner,
  Phys.\ Lett.\ B {\bf 257}, 83 (1991).



 

\bibitem{Haber:1990aw}
 H.~E.~Haber and R.~Hempfling,
  Phys.\ Rev.\ Lett.\  {\bf 66}, 1815 (1991).


\bibitem{Okada:1990gg} 
  Y.~Okada, M.~Yamaguchi and T.~Yanagida,
  Phys.\ Lett.\ B {\bf 262}, 54 (1991).



\bibitem{Ellis:1991zd} 
  J.~R.~Ellis, G.~Ridolfi and F.~Zwirner,
  Phys.\ Lett.\ B {\bf 262}, 477 (1991).

  
\bibitem{Kawasaki:2008qe} 
  M.~Kawasaki, K.~Kohri, T.~Moroi and A.~Yotsuyanagi,
  Phys.\ Rev.\ D {\bf 78}, 065011 (2008)
  [arXiv:0804.3745 [hep-ph]].



\bibitem{Yamaguchi:2016oqz} 
  M.~Yamaguchi and W.~Yin,
  doi:10.1093/ptep/pty002
  arXiv:1606.04953 [hep-ph].
 

 \bibitem{fc} 
  O.~Bruning, O.~Dominguez, S.~Myers, L.~Rossi, E.~Todesco and F.~Zimmermann,
  arXiv:1108.1617 [physics.acc-ph].
    \bibitem{fc1} 
  FCC Collaboration, 
 Tech. Rep.FCC-ACC-SPC-0001, CERN, 2014\\
    \bibitem{fc2}
   CEPC-SPPC Study Group,
  http://cepc.ihep.ac.cn/preCDR/main\_preCDR.pdf, 2015\\

    \bibitem{fc3}
  H.~Abramowicz {\it et al.},
  arXiv:1608.07538 [hep-ex]\\

    \bibitem{fc4}
  Y.~Alexahin {\it et al.},
  arXiv:1308.2143 [hep-ph]\\

  
  
\bibitem{Patrignani:2016xqp} 
  C.~Patrignani {\it et al.} [Particle Data Group],
  Chin.\ Phys.\ C {\bf 40}, no. 10, 100001 (2016).

  
  
  
\bibitem{Aoki:2016frl} 
  S.~Aoki {\it et al.},
  Eur.\ Phys.\ J.\ C {\bf 77}, no. 2, 112 (2017)
  [arXiv:1607.00299 [hep-lat]].


\bibitem{Jang:2017ieg} 
  Y.~C.~Jang {\it et al.} [SWME Collaboration],
  arXiv:1710.06614 [hep-lat].
  
\bibitem{Rattazzi:2003rj} 
  R.~Rattazzi, C.~A.~Scrucca and A.~Strumia,
  Nucl.\ Phys.\ B {\bf 674}, 171 (2003)
  [hep-th/0305184].
  
\bibitem{Gabbiani:1996hi} 
  F.~Gabbiani, E.~Gabrielli, A.~Masiero and L.~Silvestrini,
  Nucl.\ Phys.\ B {\bf 477}, 321 (1996)
  [hep-ph/9604387].
  
  
  
  
  
\bibitem{Altmannshofer:2009ne} 
  W.~Altmannshofer, A.~J.~Buras, S.~Gori, P.~Paradisi and D.~M.~Straub,
  Nucl.\ Phys.\ B {\bf 830}, 17 (2010)
  [arXiv:0909.1333 [hep-ph]].

  \bibitem{Kugo:2010fs} 
  T.~Kugo and T.~T.~Yanagida,
  Prog.\ Theor.\ Phys.\  {\bf 124}, 555 (2010)
  [arXiv:1003.5985 [hep-th]].

  
\bibitem{Goto:1990me} 
  T.~Goto and T.~Yanagida,
  Prog.\ Theor.\ Phys.\  {\bf 83}, 1076 (1990).
%


  
  \bibitem{Djouadi:2002ze} 
  A.~Djouadi, J.~L.~Kneur and G.~Moultaka,
  Comput.\ Phys.\ Commun.\  {\bf 176}, 426 (2007)
  [hep-ph/0211331].




\bibitem{Chigusa:2016ody} 
  S.~Chigusa and T.~Moroi,
  Phys.\ Rev.\ D {\bf 94}, no. 3, 035016 (2016)
  [arXiv:1604.02156 [hep-ph]].

\bibitem{Chigusa:2017drd} 
  S.~Chigusa and T.~Moroi,
  PTEP {\bf 2017}, no. 6, 063B05 (2017)
  [arXiv:1702.00790 [hep-ph]].
  
  
  
\bibitem{Buttazzo:2013uya} 
  D.~Buttazzo, G.~Degrassi, P.~P.~Giardino, G.~F.~Giudice, F.~Sala, A.~Salvio and A.~Strumia,
  JHEP {\bf 1312}, 089 (2013)
  [arXiv:1307.3536 [hep-ph]].
  
\bibitem{Giudice:2011cg} 
  G.~F.~Giudice and A.~Strumia,
  Nucl.\ Phys.\ B {\bf 858}, 63 (2012)
  [arXiv:1108.6077 [hep-ph]].
 
\bibitem{Pierce:1996zz} 
  D.~M.~Pierce, J.~A.~Bagger, K.~T.~Matchev and R.~j.~Zhang,
  Nucl.\ Phys.\ B {\bf 491}, 3 (1997)
  [hep-ph/9606211].

   
\bibitem{Vega:2015fna} 
  J.~Pardo Vega and G.~Villadoro,
  JHEP {\bf 1507}, 159 (2015)
  [arXiv:1504.05200 [hep-ph]].
 
   \bibitem{feynhiggs}
  S.~Heinemeyer, W.~Hollik and G.~Weiglein,
  Comput.\ Phys.\ Commun.\  {\bf 124}, 76 (2000)
  [hep-ph/9812320].
  
  \bibitem{feynhiggs2}
  S.~Heinemeyer, W.~Hollik and G.~Weiglein,
  Eur.\ Phys.\ J.\ C {\bf 9}, 343 (1999)
  [hep-ph/9812472].
  \bibitem{feynhiggs3}
  G.~Degrassi, S.~Heinemeyer, W.~Hollik, P.~Slavich and G.~Weiglein,
  Eur.\ Phys.\ J.\ C {\bf 28}, 133 (2003)
  [hep-ph/0212020];
  \bibitem{feynhiggs4}
  M.~Frank, T.~Hahn, S.~Heinemeyer, W.~Hollik, H.~Rzehak and G.~Weiglein,
  JHEP {\bf 0702}, 047 (2007)
  [hep-ph/0611326].
  \bibitem{feynhiggs5}
  T.~Hahn, S.~Heinemeyer, W.~Hollik, H.~Rzehak and G.~Weiglein,
  Phys.\ Rev.\ Lett.\  {\bf 112}, no. 14, 141801 (2014)
  [arXiv:1312.4937 [hep-ph]].

\bibitem{feynhiggs6} 
  H.~Bahl and W.~Hollik,
  Eur.\ Phys.\ J.\ C {\bf 76}, no. 9, 499 (2016)
  [arXiv:1608.01880 [hep-ph]].
  
\bibitem{feynhiggs7} 
  H.~Bahl, S.~Heinemeyer, W.~Hollik and G.~Weiglein,
  Eur.\ Phys.\ J.\ C {\bf 78}, no. 1, 57 (2018)
  [arXiv:1706.00346 [hep-ph]].

\bibitem{Athron:2016fuq}
P.~Athron, J.~h.~Park, T.~Steudtner, D.~St\"{o}ckinger and A.~Voigt,
  JHEP {\bf 1701}, 079 (2017)
  [arXiv:1609.00371 [hep-ph]].
\bibitem{Athron:2017fvs} 
  P.~Athron, M.~Bach, D.~Harries, T.~Kwasnitza, J.~h.~Park, D.~St\"{o}ckinger, A.~Voigt and J.~Ziebell,
  arXiv:1710.03760 [hep-ph].



\bibitem{Draper:2016pys} 
  P.~Draper and H.~Rzehak,
  Phys.\ Rept.\  {\bf 619}, 1 (2016)
  [arXiv:1601.01890 [hep-ph]].





\bibitem{Bagnaschi:2017xid} 
  E.~Bagnaschi, J.~Pardo Vega and P.~Slavich,
  Eur.\ Phys.\ J.\ C {\bf 77}, no. 5, 334 (2017)
  [arXiv:1703.08166 [hep-ph]].




  
  \bibitem{Aaboud:2017vwy}
  M.~Aaboud {\it et al.} [ATLAS Collaboration],
  arXiv:1712.02332 [hep-ex].
  
 
\bibitem{Cohen:2013xda} 
  T.~Cohen, T.~Golling, M.~Hance, A.~Henrichs, K.~Howe, J.~Loyal, S.~Padhi and J.~G.~Wacker,
  JHEP {\bf 1404}, 117 (2014)
  [arXiv:1311.6480 [hep-ph]].

\bibitem{Arkani-Hamed:2015vfh} 
  N.~Arkani-Hamed, T.~Han, M.~Mangano and L.~T.~Wang,
  Phys.\ Rept.\  {\bf 652}, 1 (2016)
  [arXiv:1511.06495 [hep-ph]].
  
  
\bibitem{Golling:2016gvc} 
  T.~Golling {\it et al.},
  CERN Yellow Report, no. 3, 441 (2017)
  [arXiv:1606.00947 [hep-ph]].

\bibitem{Hisano:2006nn} 
  J.~Hisano, S.~Matsumoto, M.~Nagai, O.~Saito and M.~Senami,
  Phys.\ Lett.\ B {\bf 646}, 34 (2007)
  [hep-ph/0610249].
\bibitem{Ellis:2015vaa} 
  J.~Ellis, F.~Luo and K.~A.~Olive,
  JHEP {\bf 1509}, 127 (2015)
  [arXiv:1503.07142 [hep-ph]].
  


\bibitem{Reinert:2017aga} 
  A.~Reinert and M.~W.~Winkler,
  arXiv:1712.00002 [astro-ph.HE].






  \bibitem{Fukugita:1986hr} 
  M.~Fukugita and T.~Yanagida,
  Phys.\ Lett.\ B {\bf 174}, 45 (1986).

\bibitem{Buchmuller:2005eh} 
  W.~Buchmuller, R.~D.~Peccei and T.~Yanagida,
  Ann.\ Rev.\ Nucl.\ Part.\ Sci.\  {\bf 55}, 311 (2005)
  [hep-ph/0502169].

\bibitem{Davidson:2008bu} 
  S.~Davidson, E.~Nardi and Y.~Nir,
  Phys.\ Rept.\  {\bf 466}, 105 (2008)
  [arXiv:0802.2962 [hep-ph]].


\bibitem{Hall:1990ac} 
  L.~J.~Hall and L.~Randall,
  Phys.\ Rev.\ Lett.\  {\bf 65}, 2939 (1990).

\bibitem{Ciuchini:1998xy} 
  M.~Ciuchini, G.~Degrassi, P.~Gambino and G.~F.~Giudice,
  Nucl.\ Phys.\ B {\bf 534}, 3 (1998)
  [hep-ph/9806308].

\bibitem{Buras:2000dm} 
  A.~J.~Buras, P.~Gambino, M.~Gorbahn, S.~Jager and L.~Silvestrini,
  Phys.\ Lett.\ B {\bf 500}, 161 (2001)
  [hep-ph/0007085].

\bibitem{DAmbrosio:2002vsn} 
  G.~D'Ambrosio, G.~F.~Giudice, G.~Isidori and A.~Strumia,
  Nucl.\ Phys.\ B {\bf 645}, 155 (2002)
  [hep-ph/0207036].

\bibitem{Paradisi:2008qh} 
  P.~Paradisi, M.~Ratz, R.~Schieren and C.~Simonetto,
  Phys.\ Lett.\ B {\bf 668}, 202 (2008)
  [arXiv:0805.3989 [hep-ph]].
  
\bibitem{Colangelo:2008qp} 
  G.~Colangelo, E.~Nikolidakis and C.~Smith,
  Eur.\ Phys.\ J.\ C {\bf 59}, 75 (2009)
  [arXiv:0807.0801 [hep-ph]].
  
  \bibitem{Freitas:2007dp} 
  A.~Freitas, E.~Gasser and U.~Haisch,
  Phys.\ Rev.\ D {\bf 76}, 014016 (2007)
  [hep-ph/0702267].

  
  

\bibitem{Goto}
T.~Goto,  http://research.kek.jp/people/tgoto/.








\bibitem{CMS:2017mkt} 
  CMS Collaboration [CMS Collaboration],
  CMS-PAS-SUS-17-009.
  
\bibitem{Sirunyan:2017cwe} 
  A.~M.~Sirunyan {\it et al.} [CMS Collaboration],
  Phys.\ Rev.\ D {\bf 96}, no. 3, 032003 (2017)
  [arXiv:1704.07781 [hep-ex]].


  
  
  
\bibitem{Kagan:1999iq} 
  A.~L.~Kagan and M.~Neubert,
  Phys.\ Rev.\ Lett.\  {\bf 83}, 4929 (1999)
  [hep-ph/9908404].
  


  \bibitem{Endo:2011mc} 
  M.~Endo, K.~Hamaguchi, S.~Iwamoto and N.~Yokozaki,
  Phys.\ Rev.\ D {\bf 84}, 075017 (2011)
  [arXiv:1108.3071 [hep-ph]].
  
  \bibitem{Moroi:2011aa} 
  T.~Moroi, R.~Sato and T.~T.~Yanagida,
  Phys.\ Lett.\ B {\bf 709}, 218 (2012)
  [arXiv:1112.3142 [hep-ph]].
  
  \bibitem{Endo:2011xq} 
  M.~Endo, K.~Hamaguchi, S.~Iwamoto and N.~Yokozaki,
  Phys.\ Rev.\ D {\bf 85}, 095012 (2012)
  [arXiv:1112.5653 [hep-ph]].
  
  \bibitem{Endo:2011gy} 
  M.~Endo, K.~Hamaguchi, S.~Iwamoto, K.~Nakayama and N.~Yokozaki,
  Phys.\ Rev.\ D {\bf 85}, 095006 (2012)
  [arXiv:1112.6412 [hep-ph]].
  
\bibitem{Nakayama:2012zc} 
  K.~Nakayama and N.~Yokozaki,
  JHEP {\bf 1211}, 158 (2012)
  [arXiv:1204.5420 [hep-ph]].
  
  
  
  \bibitem{Sato:2012bf} 
  R.~Sato, K.~Tobioka and N.~Yokozaki,
  Phys.\ Lett.\ B {\bf 716}, 441 (2012)
  [arXiv:1208.2630 [hep-ph]].
  
  
\bibitem{Shimizu:2015ara} 
  Y.~Shimizu and W.~Yin,
  Phys.\ Lett.\ B {\bf 754}, 118 (2016)
  [arXiv:1509.04933 [hep-ph]].
  
  
  
\bibitem{Yin:2016pkz} 
  W.~Yin,
  Chin.\ Phys.\ C {\bf 42}, no. 1, 013104 (2018)
  [arXiv:1609.03527 [hep-ph]].
  

 
\bibitem{Higaki:2016yeh} 
  T.~Higaki, M.~Nishida and N.~Takeda,
  PTEP {\bf 2017}, no. 8, 083B04 (2017)
  [arXiv:1611.04322 [hep-ph]].
 
\bibitem{Fukuyama:2016mqb} 
  T.~Fukuyama, N.~Okada and H.~M.~Tran,
  Phys.\ Lett.\ B {\bf 767}, 295 (2017)
  [arXiv:1611.08341 [hep-ph]].



  
  
\bibitem{Endo:2013lva} 
  M.~Endo, K.~Hamaguchi, T.~Kitahara and T.~Yoshinaga,
  JHEP {\bf 1311}, 013 (2013)
  [arXiv:1309.3065 [hep-ph]].


  \bibitem{Lopez:1993vi} 
  J.~L.~Lopez, D.~V.~Nanopoulos and X.~Wang,
  Phys.\ Rev.\ D {\bf 49}, 366 (1994)
  [hep-ph/9308336].

\bibitem{Chattopadhyay:1995ae} 
  U.~Chattopadhyay and P.~Nath,
  Phys.\ Rev.\ D {\bf 53}, 1648 (1996)
  [hep-ph/9507386].


\bibitem{Moroi:1995yh} 
  T.~Moroi,
  Phys.\ Rev.\ D {\bf 53}, 6565 (1996)
  Erratum: [Phys.\ Rev.\ D {\bf 56}, 4424 (1997)]
  [hep-ph/9512396].
   
  
  
\bibitem{Marchetti:2008hw} 
  S.~Marchetti, S.~Mertens, U.~Nierste and D.~Stockinger,
  Phys.\ Rev.\ D {\bf 79}, 013010 (2009)
  [arXiv:0808.1530 [hep-ph]].

\bibitem{Degrassi:1998es} 
  G.~Degrassi and G.~F.~Giudice,
  Phys.\ Rev.\ D {\bf 58}, 053007 (1998)
  [hep-ph/9803384].
  
  

 
 
 
 
 
  \bibitem{Hagiwara:2011af} 
  K.~Hagiwara, R.~Liao, A.~D.~Martin, D.~Nomura and T.~Teubner,
  J.\ Phys.\ G {\bf 38}, 085003 (2011)
  [arXiv:1105.3149 [hep-ph]].
  
    \bibitem{Davier:2010nc} 
  M.~Davier, A.~Hoecker, B.~Malaescu and Z.~Zhang,
  Eur.\ Phys.\ J.\ C {\bf 71}, 1515 (2011)
  Erratum: [Eur.\ Phys.\ J.\ C {\bf 72}, 1874 (2012)]
  [arXiv:1010.4180 [hep-ph]].
  

  
  
  
\bibitem{Bennett:2006fi} 
  G.~W.~Bennett {\it et al.} [Muon g-2 Collaboration],
  Phys.\ Rev.\ D {\bf 73}, 072003 (2006)
  [hep-ex/0602035].




\bibitem{Roberts:2010cj} 
  B.~L.~Roberts,
  Chin.\ Phys.\ C {\bf 34}, 741 (2010)
  [arXiv:1001.2898 [hep-ex]].
  
  
%

\bibitem{Fukuda:2017jmk} 
  H.~Fukuda, N.~Nagata, H.~Otono and S.~Shirai,
  arXiv:1703.09675 [hep-ph].
  
  
\bibitem{Asai:2007sw} 
  S.~Asai, T.~Moroi, K.~Nishihara and T.~T.~Yanagida,
  Phys.\ Lett.\ B {\bf 653}, 81 (2007)
  [arXiv:0705.3086 [hep-ph]].

\bibitem{Asai:2008sk} 
  S.~Asai, T.~Moroi and T.~T.~Yanagida,
  Phys.\ Lett.\ B {\bf 664}, 185 (2008)
  [arXiv:0802.3725 [hep-ph]].


  

  

  
  
   \bibitem{Kofman:1994rk} 
 L.~Kofman, A.~D.~Linde and A.~A.~Starobinsky,
 Phys.\ Rev.\ Lett.\  {\bf 73}, 3195 (1994)
 [hep-th/9405187].

\bibitem{Kofman:1997yn} 
 L.~Kofman, A.~D.~Linde and A.~A.~Starobinsky,
 Phys.\ Rev.\ D {\bf 56}, 3258 (1997)
 [hep-ph/9704452].

\bibitem{Lerner:2009xg} 
 R.~N.~Lerner and J.~McDonald,
 Phys.\ Rev.\ D {\bf 80}, 123507 (2009)
 PhysRevD.80.123507
 [arXiv:0909.0520 [hep-ph]].

\bibitem{Okada:2010jd} 
 N.~Okada and Q.~Shafi,
 Phys.\ Rev.\ D {\bf 84}, 043533 (2011)
 [arXiv:1007.1672 [hep-ph]].

\bibitem{Bastero-Gil:2015lga} 
 M.~Bastero-Gil, R.~Cerezo and J.~G.~Rosa,
 Phys.\ Rev.\ D {\bf 93}, no. 10, 103531 (2016)
 [arXiv:1501.05539 [hep-ph]].

\bibitem{Khoze:2013uia} 
 V.~V.~Khoze,
 JHEP {\bf 1311}, 215 (2013)
 [arXiv:1308.6338 [hep-ph]].


\bibitem{Nakayama:2010kt} 
 K.~Nakayama and F.~Takahashi,
 JCAP {\bf 1011}, 009 (2010)
 [arXiv:1008.2956 [hep-ph]].
 

\bibitem{Daido:2017wwb} 
  R.~Daido, F.~Takahashi and W.~Yin,
  JCAP {\bf 1705}, no. 05, 044 (2017)
  [arXiv:1702.03284 [hep-ph]].
  
  
\bibitem{Daido:2017tbr} 
  R.~Daido, F.~Takahashi and W.~Yin,
  doi:10.1007/JHEP02(2018)104
  arXiv:1710.11107 [hep-ph].

  
    
  \bibitem{Harigaya:2015iva} 
  K.~Harigaya, T.~T.~Yanagida and N.~Yokozaki,
  PTEP {\bf 2015}, no. 8, 083B03 (2015)
  [arXiv:1504.02266 [hep-ph]].

  


  
  
\bibitem{llstau} 
  J.~Heisig and J.~Kersten,
  Phys.\ Rev.\ D {\bf 84}, 115009 (2011)
  [arXiv:1106.0764 [hep-ph]].
 
  \bibitem{llstau2}
  J.~Heisig and J.~Kersten,
  Phys.\ Rev.\ D {\bf 86}, 055020 (2012)
  [arXiv:1203.1581 [hep-ph]].
    \bibitem{llstau3}
  J.~L.~Feng, S.~Iwamoto, Y.~Shadmi and S.~Tarem,
  JHEP {\bf 1512},166 (2015)
  [arXiv:1505.02996 [hep-ph]].
  
  
  
\end{thebibliography}
\end{document}